\documentclass[5p,times,sort&compress]{elsarticle}

\usepackage{xcolor}
\usepackage{lineno}
\modulolinenumbers[5]
\usepackage{hyperref}

\usepackage{amsmath}
\usepackage[ruled,vlined]{algorithm2e}
\newcommand{\yq}[1]{\textcolor{black}{#1}} %
\usepackage{fixltx2e}
\usepackage{multirow}
\usepackage{comment}

\journal{Energy Reports (Accepted on March 28, 2021)}

\begin{document}

\begin{frontmatter}

\title{\yq{Distributed Energy Trading Management for Renewable Prosumers with HVAC and Energy Storage}}


\author[A]{Qing~Yang}
\ead{yang.qing@szu.edu.cn}
\author[B]{Hao~Wang\corref{mycorrespondingauthor}}
\cortext[mycorrespondingauthor]{Corresponding author.}
\ead{hao.wang2@monash.edu}

\address[A]{College of Electronics and Information Engineering (CEI), Shenzhen University, Shenzhen, Guangdong Province, China}
\address[B]{Department of Data Science and Artificial Intelligence, Faculty of Information Technology, Monash University, Melbourne, VIC 3800, Australia}

\begin{abstract}
Heating, ventilating, and air-conditioning (HVAC) systems consume a large amount of energy in residential houses and buildings. Effective energy management of HVAC is a cost-effective way to improve energy efficiency and reduce the energy cost of residential users. \yq{This work develops a novel distributed method for the residential transactive energy system that enables multiple users to interactively optimize their energy management of HVAC systems and behind-the-meter batteries. Specifically, this method effectively reduces the cost of smart homes by employing energy trading among users to leverage their power usage flexibility without compromising the users' privacy. To achieve this goal, we design a distributed optimization algorithm based on the alternating direction method of multipliers (ADMM) to automatically operate the HVAC system and batteries, which minimizes the energy costs of users. Specifically, we decouple the optimization problem into a primal subproblem and a dual subproblem. The primal subproblem is solved by the users, and the dual subproblem is solved by the grid operator. Unlike the existing centralized method, our approach only uses the users' private information locally for solving the primal subproblem hence preserves the users' privacy. Using real-world data, we validate our proposed algorithm through extensive simulations in Matlab. The results demonstrate that our method effectively incentivizes the energy trading among the users to reduce users' peak load and reduce the overall energy cost of the system by 23\% on average.}
\end{abstract}

\begin{keyword}
Transactive energy; energy trading; smart grid; smart home; heating, ventilation, and air-conditioning (HVAC); distributed optimization
\end{keyword}

\end{frontmatter}


\section{Introduction}\label{sec:intro}

\subsection{Background and motivations}

Energy-intensive appliances consume a tremendous amount of energy and pose challenges to our electric power system, hindering the energy transition toward affordable cost and low emission \citep{TUBALLA2016710}. Among the power consumption of modern city buildings, the heating, ventilation, and air-conditioning (HVAC) system occupies a major chunk in both residential and commercial sectors \citep{pnnl, lu2012evaluation}. As an indispensable component of residential life, the HVAC system incurs a persistent and uncurtailable energy cost in the users' homes. Fortunately, recent innovations of smart grid technology inspire various methods that can improve the efficiency and reduce the cost of \yq{the} traditional HVAC system.

\begin{figure*}[!ht]
    \centering
    \includegraphics[width=17cm]{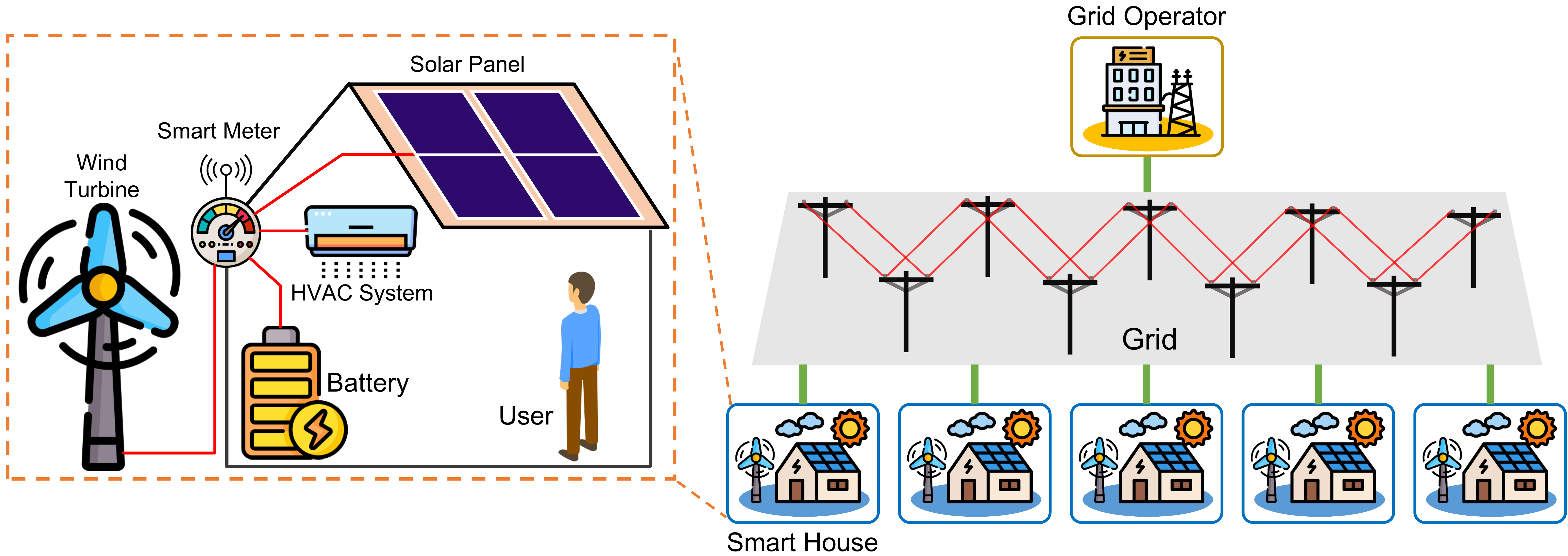}
    \caption{The system model of the distributed HVAC management with renewable generators and batteries. The smart home consists of solar/wind power generators, an HVAC system, a battery, and a smart meter. Multiple smart homes are connected through the local grid.}
    \label{f1:sysmod}
\end{figure*}

The widely installed renewable energy generators (e.g., photovoltaics and wind turbines) witness the increasing application of distributed energy resources (DER). As shown in Fig. \ref{f1:sysmod}, the local solar panel and wind turbine generate green energy that can be used to support the HVAC system. Furthermore, the microgrid allows the residential users to exchange their surplus energy with other users via the peer-to-peer (P2P) energy trading \citep{wang2016incentivizing}. Finally, modern smart meters and home energy management systems can automatically manage and control the appliances to optimize electricity usage \citep{tushar2014three}. Recently, with the popularization of the battery energy storage, smart homes can store renewable energy at a low cost using behind-the-meter lithium-ion batteries (e.g., the Tesla Powerwall), which offers more flexibility to address the above challenges. This work develops a distributed energy management method for HVAC to maximize the efficiency of DERs with optimal usage scheduling and P2P energy trading.

\subsection{Related works}
The energy optimization of HVAC emerges as an important topic in both the research community and industry of smart grids. Yu et~al. \cite{yu2017distributed} proposed a real-time automatic control method that utilizes the Lyapunov optimization tool to minimize the total cost of the HVAC system. Other studies in \cite{TRCKA201093, ALIBABAEI201681} suggested a holistic simulation model for the HVAC system and Keshtkar et~al. \cite{KESHTKAR201768} designed a supervised fuzzy learning method for HVAC to meet the users' preferences with minimal cost. Kusiak et~al. \cite{kusiak2010modeling} introduced a data-driven modeling method to achieve optimal in-home HVAC load scheduling. \yq{The authors in \citep{mohamed2020multi} studied the energy management of an energy hub, a networked multi-microgrid with three agents, and a transportation system.}
Song et~al. \cite{song2020energy} studied energy efficiency of end-user groups in multi-zone buildings, and the findings could help personalize appropriate control strategies for HVAC systems.
Fong et~al. \cite{fong2009system} designed an evolutionary optimization algorithm for load reduction of a centralized HVAC system. Nguyen et~al. \cite{nguyen2014energy} considered solar energy for HVAC and developed a stochastic optimization method to minimize the cost. Reddy et~al. \cite{reddy2020exergy} developed an exergy-based model predictive controller for a micro-scale concentrated solar power system to minimize the energy consumption of building HVAC systems.
Wu and Skye \cite{wu2018net} studied HVAC technologies paired with a PV system to achieve net-zero energy for residential buildings in different climate zones of $15$ representative cities across the US. \yq{Gong et~al. developed a primal-dual method of multipliers approach to solve the operation management of a smart energy hub and a microgrid in clean smart islands \citep{gong2020towards}. In our work, we focus on a different scenario, where smart homes interact with each other to trade energy and schedule their HVAC units and energy storage.}

\yq{In \cite{wang2015bargaining}, the authors designed a distributed energy trading algorithm based on the primal-dual method for interconnected microgrids. However, the standard primal-dual algorithm may have a slow convergence problem, especially when solving optimization problems with linear objective functions. The alternating direction method of multipliers (ADMM) algorithm significantly improves the convergence and has been reported in studies on energy trading \cite{lee2019optimal}. Our work adopts it to derive a distributed energy trading strategy for smart homes. However, ADMM also has drawbacks, e.g., the convergence cannot always be guaranteed for multi-block iteration, even the optimization problem is convex. But our design only involves two blocks, and thus the convergence is ensured. The consensus algorithm is another paradigm for a decentralized algorithm. Instead of having a common node to handle the multipliers, the consensus algorithm lets all the nodes remain a copy of multipliers and exchange information with other peer nodes (e.g., neighbor nodes) to reach the consensus of the multipliers for all the nodes. Consensus algorithm has been adopted to design decentralized energy management for both generation and demand sides in smart grid \cite{zhao2016consensus} and solve the economic dispatch problem in power systems \cite{yang2013consensus}.}

Unlike the above literature that only considered the standalone HVAC management within a single house or apartment, several recent studies investigated cooperative HVAC management across multiple dwellings in the microgrid. Luna et~al. \cite{luna2016cooperative} proposed a mixed-integer linear programming method to solve the cooperative energy scheduling problem for multiple smart homes. A near-optimal P2P energy trading algorithm is proposed by Alam et~al. \cite{ALAM20191434} to reduce the electricity cost of all participating smart homes. However, both these two works employed a centralized optimization method that relies on a central party to collect users' information and make decisions, which causes privacy concerns since all the users have to reveal their electricity usage information to this central party. As pointed out in \cite{pnnl}, the residential electricity usage records contain rich privacy information of the users. Various methods that can exploit privacy information from users' power usage records were discussed in \cite{eibl2014influence}. Eibl et~al. \cite{eibl2014influence} discussed the tradeoff between the performance of HVAC systems and the leakage of the users' privacy. By contrast, we develop a distributed HVAC management algorithm that can protect the users' privacy and minimize the overall cost simultaneously. \yq{Instead of modeling the general load of prosumers or microgrids, our work identifies a more realistic energy trading scenario. The use of HVAC exhibits diverse patterns as different households have different occupancy patterns.} Our preliminary work in \cite{yang2020cooperative} explored the cooperative energy management of HVAC units via distributed optimization. \yq{In this work, we further consider energy storage, which adds more flexibility into the cooperative energy management and magnifies the benefit of energy trading.}

As the cost of lithium-ion batteries decreases, a high-capacity battery energy storage system (BESS) becomes available and affordable for residential users. For example, the Tesla Powerwall offers a capacity of 13.5kWh and 7kW peak discharging output power \cite{powerwall}.   Rahimi-Eichi et~al. \cite{rahimi2013battery} surveyed the challenges and solutions of applying BESS in smart grids. Lucas et~al. \cite{LUCAS201626} proposed to employ a high-capacity vanadium redox flow battery for frequency regulation and peak shaving. Such et~al. \cite{such2012battery} presented several control methods that integrate BESS with wind-power generators to address the variability of wind generation. In this work, we adopt the BESS in the smart homes and jointly manage the battery and the HVAC to achieve optimal energy utilization. \yq{Our work aims to evaluate the benefit of cooperative energy management of HVAC systems with energy storage to enable a more applicative strategy for the energy trading toward its wide implementation.}

\subsection{Novelty and contributions}
This work presents a distributed energy management method for HVAC systems that incorporates P2P energy trading and rechargeable battery to reduce the overall cost while preserving the users' privacy. We summarize the contributions of this work as follows.

\begin{enumerate}
  \item We integrate the latest smart grid technologies of P2P energy trading and BESS to optimize the energy management of HVAC systems.
  \item We design a distributed optimization algorithm based on ADMM to automatically manage the HVAC units and BESS, which minimizes the users' costs without leaking their private information.
  \item We validate our proposed method by extensive simulations using real-world user data, and the results show an average cost reduction of 23\%.
\end{enumerate}

Outline of the manuscript: Section~\ref{sec:model} introduces the HVAC energy management model on the transactive energy platform. Section~\ref{sec:formulation} formulates an optimization problem for the cooperative energy system. Section~\ref{sec:solution} elaborates on the design of the distributed P2P energy trading algorithm for HVAC. Section~\ref{sec:eval} evaluates the proposed P2P energy trading system with extensive simulations and Section~\ref{sec:conclusion} concludes our work.

\section{System model}\label{sec:model}

\begin{figure*}[!ht]
    \centering
    \includegraphics[width=15cm]{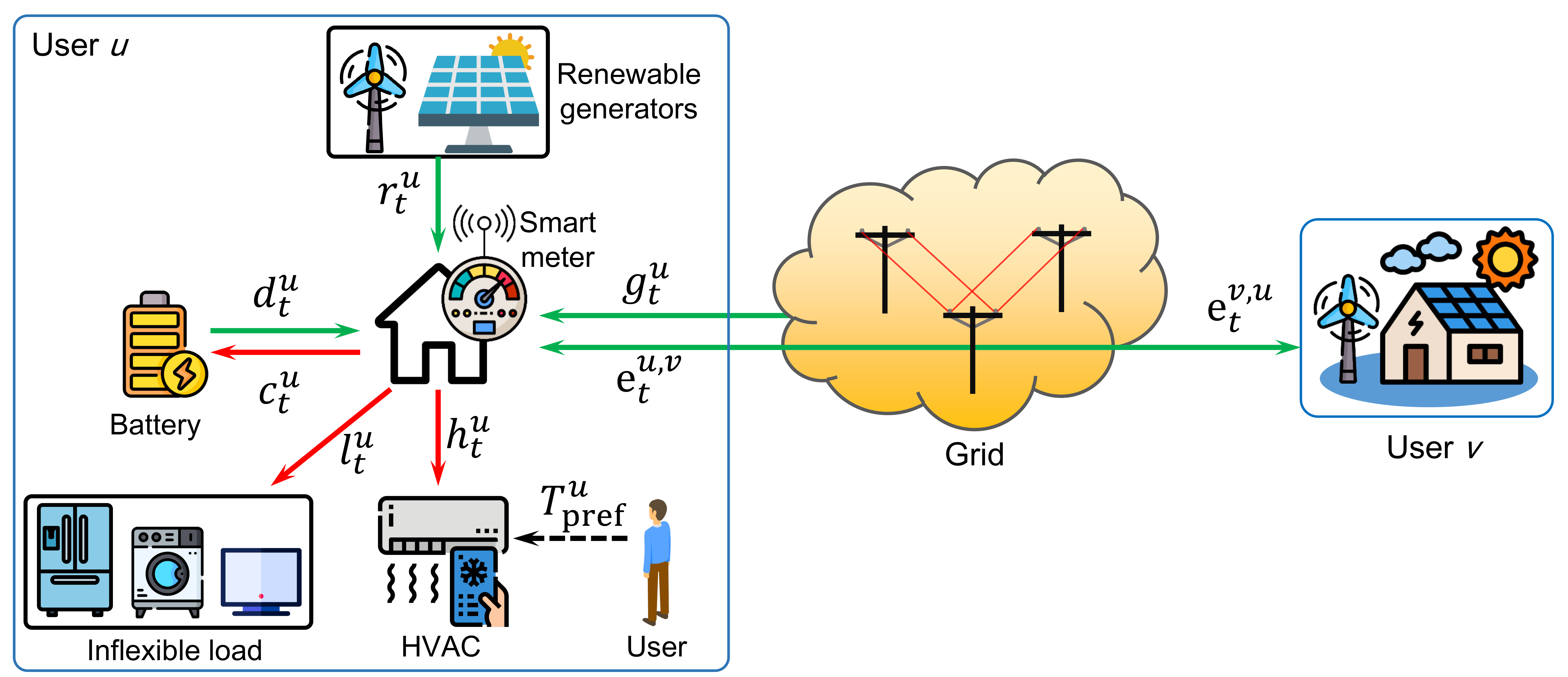}
    \caption{The operation of the HVAC system.}
    \label{f2:prin}
\end{figure*}

\begin{table*}[!t]
    \centering
    \yq{\caption{Nomenclature}\label{tab:def}
    \footnotesize
    \renewcommand{\arraystretch}{1.1}
    \begin{tabular}{|l l|}
        \hline
        \textbf{Notation} & \textbf{Definition} \\
        \hline
        $N$ & The total number of users \\
        $H$ & The total number of time slots of the operation time \\
        $\mathbf{U} {=} \{1,..,N\}$ & The set of users \\
        $u \in \mathbf{U}$ & The residential user \\
        $\mathbf{H} {=} \{1,..,H\}$ & The system's total operation time \\
        $g_{t}^{u}$ & the amount of electricity that user $u$ purchases from the grid in the $t$-th time slot \\
        $G^{u}$ & The maximal power that user $u$ can draw from the grid \\
        $\mathcal{C}^u_{\mathrm{grid}}$ & The grid electricity payment of user $u$ \\
        $\gamma_{\mathrm{g}}$ & The normal price of the grid electricity \\
        $\gamma_{\mathrm{p}}$ & The peak price of the grid electricity \\
        $r_{t}^{u}$ & User $u$'s renewable energy usage in the $t$-th time slot \\
        $R_{t}^{u}$ &  The maximal renewable energy that user $u$ can use in $t$-th time slot \\
        $e_{t}^{u,v}$ & The amount of energy that user $u$ purchased from user $v$ in the $t$-th time slot \\
        $B^{u}$ & The capacity of the user $u$'s battery \\
        $P^{u}_{\mathrm{cha}}$ & The maximal charging power of user $u$ per time slot \\
        $P^{u}_{\mathrm{dis}}$ & The maximal discharging power of user $u$ per time slot \\
        $b_{t}^u$ & The energy stored in the battery of user $u$ in the $t$-th time slot \\
        $c_{t}^u$ & The energy charged into user $u$'s battery in the $t$-th time slot \\
        $d_{t}^u$ & The energy discharged from user $u$'s battery in the $t$-th time slot \\
        $R$ and $C$ & The equivalent heat capacity and thermal resistance of the HVAC system \\
        $\alpha^u$ & The working mode and efficiency of user $u$'s HVAC system \\
        $\mathcal{C}^u_{\mathrm{battery}}$ & User $u$'s cost of the battery \\
        $\tau^{u}_{t}$ & User $u$'s  indoor temperature \\
        $T^{u}_{\mathrm{pref}}$ & User $u$'s preferred indoor temperature \\
        $T_{t}$ & The outdoor environmental temperature in the $t$-th time slot \\
        $h^{u}_t$ & The energy consumed by user $u$'s HVAC system in the $t$-th time slot \\
        $\mathcal{C}^{u}_{\mathrm{hvac}}$ & User $u$'s discomfort cost of the HVAC system \\
        $l^{u}_{t}$ & User $u$'s inflexible load in the $t$-th time slot \\
        \hline
    \end{tabular}}
\end{table*}

We illustrate the system model of the distributed HVAC management system in Fig.~\ref{f1:sysmod}. As shown in the left part of the figure, the smart home is equipped with renewable generators (e.g., wind turbines and rooftop solar panels) that can generate renewable energy to power appliances. The HVAC system \yq{consists of the} air conditioner, heater, and ventilation equipment, which can keep the indoor temperature at the user's preference. A behind-the-meter battery is installed in the home to store the extra electric energy for future use. The smart meter is an IoT device that can run scheduling algorithms to coordinate the above appliances. In addition, smart homes can also exchange information with others with smart grid communication technologies such as 5G-Narrowband and LoRa \citep{ma2013smart}.

In this work, we consider a local grid that consists of multiple smart homes 
as shown in the right part of Fig.~\ref{f1:sysmod}. We denote the residential users (with their smart homes) in the grid by a set $\mathbf{U} {=} \{1,..,N\}$, where $N$ is the total number of users. We assume that the user can manage its in-home appliances and trade energy with other users in the unit of \emph{time slot}, which represents the minimal operational time interval of the system. Without loss of generality, we assume that the time slot is one hour and the total operation time is $\mathbf{H} {=} \{1,..,H\}$. \yq{We assume the day-ahead scheduling as described in \citep{nan2018optimal} throughout this work. Therefore the range of the operation time $H$ is $24$. The definition of the variables used in this paper is listed in Tab.~\ref{tab:def}.}

\subsection{The power supply of the smart home}

For each user $u \in \mathbf{U}$, its power supply consists of three sources: (1) the user can use the electricity from the grid at a price that is set by the grid operator; (2) the user can also generate green energy via its renewable generators; (3) the user can buy electricity from other users in the grid with the P2P energy trading. As shown in Fig.~\ref{f2:prin}, we denote $g_{t}^{u}$ as the amount of electricity that user $u$ purchases from the grid in the $t$-th time slot, then we have the following constraint:
    \begin{equation}
        0 \leq  g_{t}^{u} \leq G^{u}, ~ \forall u \in \mathbf{U}, \forall t \in \mathbf{H}, \label{constraint-load6}
    \end{equation}
where $G^{u}$ denotes the maximal power that user $u$ can draw from the grid, which is limited by its line capacity. The grid operator adopts the two-part tariff (TPT) pricing plan \citep{bustos2019evolution} with a fixed energy price $\gamma_{\mathrm{g}}$ and a peak price $\gamma_{\mathrm{p}}$. Therefore, the grid electricity payment of user $u$ is written as
    \begin{align}
            \mathcal{C}^u_{\mathrm{grid}} = \gamma_{\mathrm{g}} \sum_{t\in \mathbf{H}} g_{t}^{u} + \gamma_{\mathrm{p}} \max_{t\in \mathbf{H}} g_{t}^{u}, \label{objective-supply} 
    \end{align}
where the first part of the payment $\gamma_{\mathrm{g}} \sum_{t\in \mathbf{H}} g_{t}^{u}$ is the energy charge and the second part $\gamma_{\mathrm{pg}} \max_{t\in \mathbf{H}} g_{t}^{u}$ is the peak load charge. The grid operator introduces the peak charge to shave the peak load of the grid.

The second source of the power supply is the renewable energy generated by the wind turbine and solar panel. We let $r_{t}^{u}$ denotes the summation of wind and solar energy generated by the user $u$'s smart home in the $t$-th time slot as shown in Fig.~\ref{f2:prin}. In practice, the value of $r_{t}^{u}$ is bounded by several factors, including the geographical location of the home, wind and solar condition, and the capacity of the renewable generators. Hence, the following constraint is established for renewable energy
    \begin{align}
        0 \leq r_{t}^{u} \leq R_{t}^{u}, ~ \forall u \in \mathbf{U}, \forall t \in \mathbf{H},  \label{constraint-load5}
    \end{align}
where $R_{t}^{u}$ is the maximal renewable energy that user $u$ can use in $t$-th time slot. In this work, we will use the real data collected from practical wind turbines and solar panels to simulate $R_{t}^{u}$.

\subsection{P2P energy trading}
The third source of the power supply is through P2P energy trading with other grid users. We adopt the P2P energy trading model as proposed in \cite{alam2019peer} and \cite{nguyen2018optimizing}, in which users can exploit their diversities of power supply and usage manner to reduce their cost of electricity. With P2P energy trading, user $u$ can trade (buy or sell) energy with another user $v$ at a fixed price, which is lower than the price of the grid. The P2P trading is mutually beneficial for both the users since the sellers can earn profits, and the buyers can obtain cheaper electricity.

The P2P energy trading market allows users to trade energy with other users in the grid. Let $e_{t}^{u,v}$ denotes the amount of energy that user $u$ purchased from user $v$ in the $t$-th time slot. Note that the value of $e_{t}^{u,v}$ can be negative, which means user $u$ sells its energy to user $v$. We omit the loss of electricity in the power transfer in the grid since the users are close to each other.
The following constraint must be satisfied 
    \begin{align}
        e_t^{u,v} &= -e_t^{v,u},~\forall t \in \mathbf{H},~\forall u,v \in \mathbf{U}, \label{constraint-trading1} \\
        e_t^{u,v} &= 0,~\textrm{if}~u = v, \label{constraint-trading2}
    \end{align}
where Eq.~\eqref{constraint-trading1} clears the trading market, and Eq.~\eqref{constraint-trading2} prohibits the user from trading with itself.

\subsection{The battery energy storage system (BESS)}

The BESS can store the surplus energy from the renewable energy generators and discharge it when needed to help build a sustainable and secure smart home \citep{yang2018battery}. Let $B^{u}$ denote the capacity of the battery in user $u$'s home, $P^{u}_{\mathrm{cha}}$ and $P^{u}_{\mathrm{dis}}$ denote the maximal charging and discharging power of the battery in each time slot, respectively. The parameters $B^{u}$, $P^{u}_{\mathrm{cha}}$, and $P^{u}_{\mathrm{dis}}$ depend on the specific BESS device, for example, the Tesla Powerwall~2 has $B^{u} = 13.5$ in kWh, $P^{u}_{\mathrm{cha}} = 7$, and $P^{u}_{\mathrm{dis}} = 7$ in kW.

To model the operation of the battery, we use $b_{t}^u$ to denote the energy stored in the battery of user $u$ in the $t$-th time slot. Let $c_{t}^u$ and  $d_{t}^u$ denote the energy charged into and discharged from the user $u$'s battery in the $t$-th time slot. Hence, the operation status of the battery can be modeled as
    \begin{equation}
        b_{t}^u = b_{t-1}^u +  c_{t}^u - d_{t}^u, \forall u \in \mathbf{U}, t \in \mathbf{H}, \label{constraint-load7} \\
    \end{equation}
with the following constraints $b_{t}^u \in [0, B^{u}]$, $c_{t}^u \in [0, P^{u}_{\mathrm{cha}}]$, and $d_{t}^u \in [0, P^{u}_{\mathrm{dis}}]$. Moreover, the charging/discharging operation also incurs degradation of the battery. We model user $u$'s battery degradation as the cost of the BESS in the following equation:
    \begin{align}
            \mathcal{C}^u_{\mathrm{battery}} = \gamma_{\mathrm{b}} \sum_{t\in \mathbf{H}} \left( c_{t}^u + d_{t}^u \right), \label{objective-battery}
    \end{align}
where $\gamma_{\mathrm{b}}$ is the cost coefficient of the battery. \yq{Note that a battery can support a certain number of full charge/discharge cycle of charging and discharge over its lifespan, and a normalized unit cost for charge and discharge can be calculated, i.e., the cost coefficient $\gamma_{\mathrm{b}}$ in our model.}
   
\subsection{Working principle of the HVAC system}

The major task of the HVAC system is to adjust the indoor temperature, denoted by $\tau^{u}_{t}$, according to user $u$'s preference $T^{u}_{\mathrm{pref}}$. We assume that the energy consumed by the HVAC system is $h^{u}_t$ in $t$-th time slot for the cooling or heating operation. In this study, we focus on optimizing the performance of the HVAC system and consider the other appliances 
as inflexible load. We let $l^{u}_{t}$ denote user $u$'s inflexible load in the $t$-th time slot.

Based on the HVAC model proposed by Lu et~al. \cite{lu2012evaluation}, the indoor temperature $\tau^{u}_{t}$ follows the dynamic equation below
    \begin{equation}
            \tau^{u}_{t+1} = T_{t+1} - \left( T_{t+1} - \tau^{u}_{t} \right) e^{1/RC} + \alpha^u h^{u}_t,~ \forall u \in \mathbf{U}, t \in \mathbf{H}, \label{constraint-load1}
    \end{equation}
where $T_{t}$ denotes the outdoor temperature in the $t$-th time slot, which is the same for all the users at the same location. The coefficients $R$ and $C$ are the equivalent heat capacity and thermal resistance of the HVAC system. The parameter $\alpha^u$ denotes the working mode and efficiency of user $u$'s HVAC system: a positive $\alpha^u$ indicates heating and negative indicates cooling.

In this study, we use the discomfort of the user to measure the performance of the HVAC system. User $u$'s discomfort is the deviation of the indoor temperature $\tau^{u}_{t}$ from the user's preferred temperature $T^{u}_{\mathrm{pref}}$. We model the discomfort cost by the following equation
    \begin{equation}
            \mathcal{C}^{u}_{\mathrm{hvac}} = \gamma_{h}\sum_{t=1}^{t=H} \left( \tau^{u}_{t} - T^{u}_{\mathrm{pref}} \right)^{2}, ~ \forall u \in \mathbf{U}, \label{objective-load1}
    \end{equation}
where the coefficient $\gamma_{h}$ indicates user $u$'s sensitivity to the temperature. In Eq.~(\ref{objective-load1}), user $u$'s indoor temperature $\tau^{u}_{t}$ must follow the practical constraint $\tau^{u}_{t} \in [\tau^{*}, \tau_{*}]$, where $\tau^{*}$ and $\tau_{*}$ denote the upper and lower limits of the HVAC system, respectively.
\yq{Note that the discomfort cost in Eq.~\eqref{objective-load1} is a quadratic function of the temperature deviation, reflecting the discomfort experienced by the users.}

\section{Optimal HVAC management with P2P energy trading}\label{sec:formulation}

We present the mathematical model of the energy management problem for HVAC systems in two scenarios. In the first scenario, we consider the standalone energy management of the HVAC system in the smart home, where each user individually schedules its own electricity usage. In the second scenario, the P2P energy trading is employed to enable \yq{the cooperative operation} of users' HVAC systems and further improve the efficiency.

\yq{To illustrate the optimal HVAC management problem based on our system model, we present a normal smart home user $u$ in Hong Kong as an example. The maximum allowable electricity current of the powerline in the smart home is 40A/220V that translates to 8.8kWh per hour; therefore the upper bound of user $u$'s grid usage is $G^u = 8.8$. The user's renewable generation is bounded by its available renewable energy $R_t^u$, which is decided by the environment as shown in Fig.~\ref{f:re}. Assume the user has a Tesla Powerwall~2 installed as the BESS, then the battery capacity is $B^{u} {=} 13.5$ and the charge/discharge power is $P^{u}_{\mathrm{cha}} {=} 7$ and $P^{u}_{\mathrm{dis}} {=} 7$ according to its specification. The user's preference on the indoor temperature is fixed, for example 25$^{\circ}C$, so $T^{u}_{\mathrm{pref}} {=} 25$. The user's HVAC management is to schedule its grid supply $g_t^u$, renewable energy $r^u_t$, HVAC load $h^u_t$, inflexible load $l^u_t$, energy trading $e_t^{u,v}$, and battery operation $c^u_t, d^u_t$ to minimize the total cost.}

To shorten the notation, we redefine the variables using the column vectors as follows:
\begin{align}
    &\textrm{Grid supply:} \quad &\mathbf{g}^{u} &\triangleq \left[ g_{1}^{u}, g_{2}^{u}, \dots, g_{H}^{u} \right]^{\top}, \\
    &\textrm{Local renewable:} \quad &\mathbf{r}^{u} &\triangleq \left[ r_{1}^{u}, r_{2}^{u}, \dots, r_{H}^{u} \right]^{\top}, \\
    &\textrm{HVAC load:} \quad &\mathbf{h}^{u} &\triangleq \left[ h_{1}^{u}, h_{2}^{u}, \dots, h_{H}^{u} \right]^{\top}, \\
    &\textrm{Inflexible load:} \quad &\mathbf{l}^{u} &\triangleq \left[ l_{1}^{u}, l_{2}^{u}, \dots, l_{H}^{u} \right]^{\top}, \\
    &\textrm{Energy trading:} \quad &\mathbf{e}^{u,v} &\triangleq \left[ e_{1}^{u,v}, e_{2}^{u,v}, \dots, e_{H}^{u} \right]^{\top},
\end{align}
\yq{where symbol ${\top}$ denotes the transpose of a matrix.} And for the battery operation, we define the following notations
\begin{align}
    &\textrm{Battery charge:} \quad &\mathbf{c}^{u} &\triangleq \left[ c_{1}^{u}, c_{2}^{u}, \dots, c_{H}^{u} \right]^{\top}, \\
    &\textrm{Battery discharge:} \quad &\mathbf{d}^{u} &\triangleq \left[ d_{1}^{u}, d_{2}^{u}, \dots, d_{H}^{u} \right]^{\top}.
\end{align}

\subsection{Scenario 1 (S1): standalone energy management of HVAC}
In the standalone mode, users independently schedule their own energy usages and the HVAC system. The objective of each user is to minimize its total cost, including the electricity bill, battery cost, and the discomfort of using the HVAC system. The battery can be used to balance the supply and demand by charging $c_{t}^{u}$ and discharging $d_{t}^{u}$, which adds flexibility to the energy management.

For each user $u$, the power supply and consumption must be balanced and the energy balance constraint for each user $u \in \mathbf{U}$ is
\begin{equation}
    \mathbf{r}^{u} + \mathbf{g}^{u} + \mathbf{d}^{u} = \mathbf{h}^{u} + \mathbf{l}^{u} + \mathbf{c}^{u}, ~ \forall u \in \mathbf{U}. \label{constraint-load11}
\end{equation}
Note here the left-hand side of Eq.~\eqref{constraint-load11} represents the total energy supply, including renewable supply $\mathbf{r}^{u}$, grid supply $\mathbf{g}^{u}$, and energy discharged from the battery $\mathbf{d}^{u}$. The right-hand side of Eq.~\eqref{constraint-load11} represents the total consumption of the HVAC load $\mathbf{h}^{u}$, inflexible load $\mathbf{l}^{u}$, and the energy charged into the battery $\mathbf{c}^{u}$.

In the standalone scenario, \yq{the users optimize their own energy schedule without interacting with any other users through energy trading.} The total operating cost of user $u$ is
\begin{align}
             &\mathcal{C}_{\mathrm{S1}}^{u}(\mathbf{g}^{u}, \mathbf{h}^{u}, \mathbf{c}^{u}, \mathbf{d}^{u}) =  \mathcal{C}^u_{\mathrm{grid}} + \mathcal{C}^u_{\mathrm{battery}} + \mathcal{C}^{u}_{\mathrm{hvac}} \nonumber \\
             &{=} \gamma_{\mathrm{g}} \sum_{t\in \mathbf{H}} g_{t}^{u} {+} \gamma_{\mathrm{p}} \max_{t\in \mathbf{H}} g_{t}^{u} {+} \gamma_{\mathrm{b}} \sum_{t\in \mathbf{H}} \left( c_{t}^u {+} d_{t}^u \right) {+} \gamma_{h}\sum_{t\in \mathbf{H}} \left( \tau^{u}_{t} {-} T^{u}_{\mathrm{pref}} \right)^{2}, \label{objective-operatingcost} 
\end{align}
where $\mathcal{C}^u_{\mathrm{grid}}$ is the payment to the grid, $\mathcal{C}^u_{\mathrm{battery}}$ denotes the operation cost of the BESS, and $\mathcal{C}^{u}_{\mathrm{hvac}}$ denotes the the user's discomfort cost when using HVAC. 

Therefore, the HVAC management of user $u$ is to minimize its total cost of Eq.~\eqref{objective-operatingcost}. This leads to the following optimization problem:
    \begin{equation}
        \begin{aligned}
            \mathbf{M}^u_{S1} = & \arg \min_{\mathbf{g}^{u}, \mathbf{h}^{u}, \mathbf{c}^{u}, \mathbf{d}^{u}, \mathbf{r}^{u}} \mathcal{C}_{\mathrm{S1}}^{u}(\mathbf{g}^{u}, \mathbf{h}^{u}, \mathbf{c}^{u}, \mathbf{d}^{u}),  \label{opt1}\\
            \mathrm{s.t.}& \: \mathrm{constraints} \:  \mathrm{\eqref{constraint-load6}}, \mathrm{\eqref{constraint-load5}}, 
            \mathrm{\eqref{constraint-load7}},\mathrm{\eqref{constraint-load1}},\mathrm{\eqref{constraint-load11}}, \\
        \end{aligned}
    \end{equation}
which solves the optimal HVAC energy schedule $\mathbf{M}^u_{S1}$ to minimize user $u$'s total cost. According to Eq.~\eqref{opt1}, user $u$ can locally solve the optimization problem with available optimization tools since it is a standard convex optimization. Hence, we omit the solution method for Eq.~\eqref{opt1} and let $\mathbf{M}^u_{S1}$ serve as a benchmark cost for the comparison with the costs in the second scenario.

\subsection{Scenario 2 (S2): HVAC management with energy trading}

\begin{figure*}[!ht]
    \centering
    \includegraphics[width=16cm]{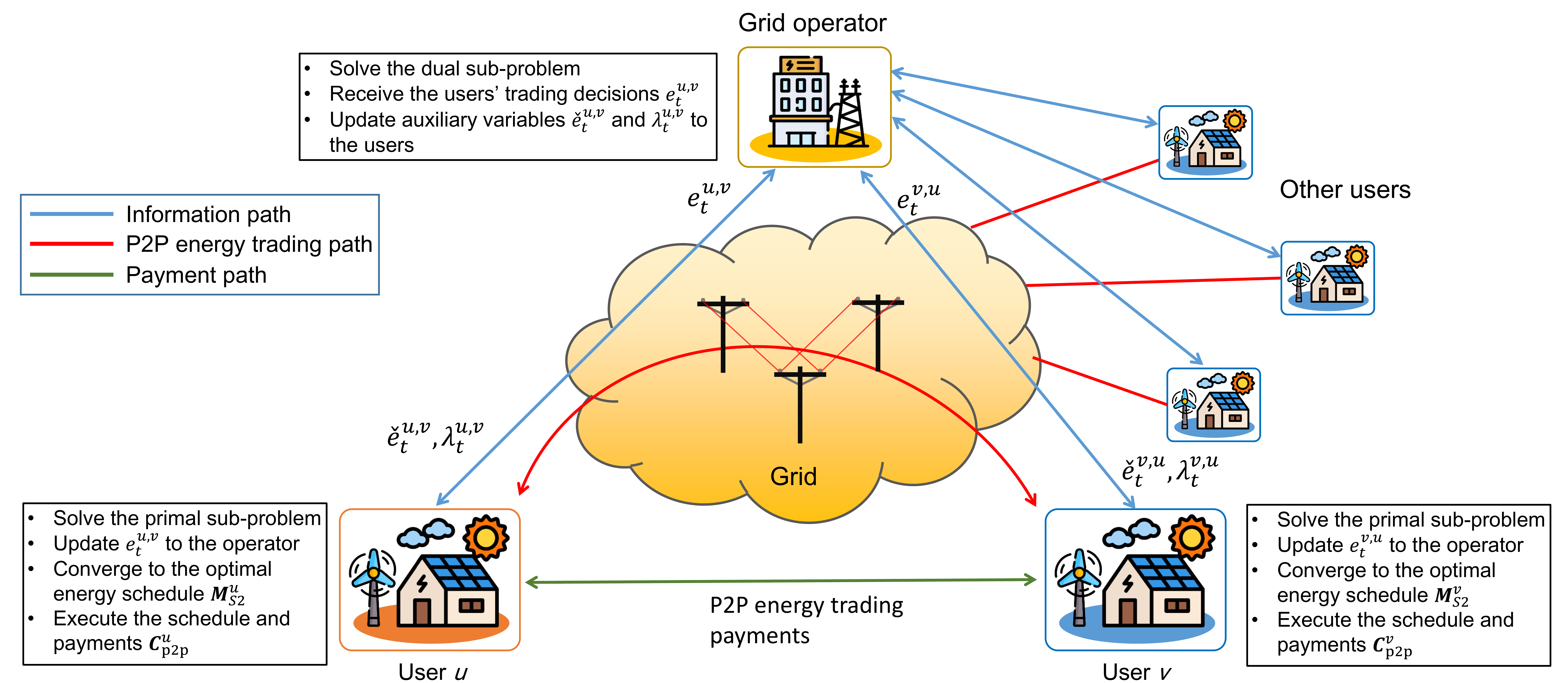}
    \caption{The distributed HVAC energy management process. Here we show the P2P energy trading between two users ($u$ and $v$) as an example.}
    \label{f3:process}
\end{figure*}

In this scenario, users not only schedule their internal energy supply and demand but also exchange energy externally with other users when needed in the cooperative scenario. Since the users can sell extra energy to other users or buy energy from other users, the constraint in Eq.~\eqref{constraint-load11} is changed to
    \begin{equation}
        \begin{aligned}
            \mathbf{r}^{u} + \mathbf{g}^{u} + \mathbf{d}^{u} + \sum_{v \in \mathbf{U}}\mathbf{e}^{u,v} = \mathbf{h}^{u} + \mathbf{l}^{u} + \mathbf{c}^{u}, ~ \forall u \in \mathbf{U}, \label{constraint-load12}
        \end{aligned}
    \end{equation}
where the term $\sum_{v \in \mathbf{U}}\mathbf{e}^{u,v}$ represents the total energy that user $u$ buys from (or sells to) other users via P2P energy trading.

During P2P energy trading, the energy buyer pays to the energy seller for the energy traded. We assume that the grid operator charges zero fees to the trading users, and sets a fixed electricity price $\phi^{\mathrm{p2p}}_t$ for the P2P energy trading. Hence, user $u$'s overall payment for the P2P energy trading is
    \begin{align}
        \mathcal{C}^{u}_{\mathrm{p2p}} = \sum_{t \in \mathbf{H}} \left( \phi^{\mathrm{p2p}}_t \sum_{v \in \mathbf{U}} e^{u,v}_{t} \right). \label{objective-tradingpayment}
    \end{align}

In this scenario, the total cost includes not only the original cost in Eq.~\eqref{objective-operatingcost}, but also the energy-trading payment. Hence, the total cost of user $u$ in scenario 2 is
\begin{align}
             &\mathcal{C}_{\mathrm{S2}}^{u}(\mathbf{g}^{u}, \mathbf{h}^{u}, \mathbf{c}^{u}, \mathbf{d}^{u}, \mathbf{e}^{u,v}) =  \mathcal{C}^u_{\mathrm{grid}} + \mathcal{C}^u_{\mathrm{battery}} + \mathcal{C}^{u}_{\mathrm{hvac}} + \mathcal{C}^{u}_{\mathrm{p2p}}, \nonumber\\
             &= \gamma_{\mathrm{g}} \sum_{t\in \mathbf{H}} g_{t}^{u} + \gamma_{\mathrm{p}} \max_{t\in \mathbf{H}} g_{t}^{u} + \gamma_{\mathrm{b}} \sum_{t\in \mathbf{H}} \left( c_{t}^u {+} d_{t}^u \right) \nonumber\\
             &\quad + \gamma_{h}\sum_{t\in \mathbf{H}} \left( \tau^{u}_{t} {-} T^{u}_{\mathrm{pref}} \right)^{2} + \sum_{t \in \mathbf{H}} \left( \phi^{\mathrm{p2p}}_t \sum_{v \in \mathbf{U}} e^{u,v}_{t} \right). \label{objective-operatingcost2} 
\end{align}
\yq{Note that in Scenario 2, users not only optimize their own energy schedule but also trade energy with other users. Therefore, in addition to the payment to the grid $\mathcal{C}^u_{\mathrm{grid}}$, energy storage operating cost $\mathcal{C}^u_{\mathrm{battery}}$, and HVAC discomfort cost $\mathcal{C}^{u}_{\mathrm{hvac}}$, P2P energy trading payment/cost $\mathcal{C}^{u}_{\mathrm{p2p}}$ is also included.}

In Scenario 2, the objective of user $u$ is to optimize the operating cost $\mathcal{C}_{\mathrm{S2}}^{u}(\mathbf{g}^{u}, \mathbf{h}^{u}, \mathbf{c}^{u}, \mathbf{d}^{u}, \mathrm{e}^{u,v})$ by scheduling the user's energy supply and demand $\{ \mathbf{g}^{u}, \mathbf{h}^{u}, \mathbf{c}^{u}, \mathbf{d}^{u} \}$ and energy trading $\{ \mathrm{e}^{u,v} \}$. We formulate this problem as the following optimization problem
    \begin{equation}
        \begin{aligned}
            \mathbf{M}^u_{S2} = & \arg \min_{\mathbf{g}^{u}, \mathbf{h}^{u}, \mathbf{c}^{u}, \mathbf{d}^{u}, \mathbf{r}^{u}, \mathbf{e}^{u,v}} \mathcal{C}_{\mathrm{S2}}^{u}(\mathbf{g}^{u}, \mathbf{h}^{u}, \mathbf{c}^{u}, \mathbf{d}^{u}, \mathbf{e}^{u,v}) ,  \label{opt2}\\
            \mathrm{s.t.}& \: \mathrm{constraints} \:  \mathrm{\eqref{constraint-load6}}, \mathrm{\eqref{constraint-load5}}, \mathrm{\eqref{constraint-trading1}}, \mathrm{\eqref{constraint-trading2}},
            \mathrm{\eqref{constraint-load7}},\mathrm{\eqref{constraint-load1}},\mathrm{\eqref{constraint-load12}}, \\
        \end{aligned}
    \end{equation}
where the solution $\mathbf{M}^u_{S2}$ is the optimal internal energy scheduling (including renewable generation $\mathbf{r}^{u}$, HVAC load $\mathbf{h}^{u}$, and battery operation $\mathbf{c}^{u}, \mathbf{d}^{u}$) and the optimal external energy trading $\mathbf{e}^{u,v}$ for all users.

\section{Distributed optimization algorithm} \label{sec:solution}

As discussed in Section \ref{sec:formulation}, the optimization problem in \eqref{opt2} cannot be solved locally by the individual user because each user's optimization problem involves other users' electricity usage information. The conventional method to solve \eqref{opt2} is to let a coordinator (e.g., the grid operator) collect all users' electricity usage information and centrally solve the optimization problem. However, the centralized method leads to serious privacy concerns since all the users have to reveal their operational parameters to the central coordinator.

To preserve the users' privacy, we propose a distributed HVAC management algorithm to solve the optimization problem \eqref{opt2}. In our method, the users cooperatively schedule their renewable energy usage, HVAC load, and battery operation through the information exchange of energy trading. This algorithm only requires the users to share their energy trading decisions without revealing any private operational information.

\subsection{The augmented Lagrangian method}
We adopt the alternating direction method of multipliers (ADMM) method \cite{boyd2011distributed} to solve the optimization problem of \eqref{opt2}. First, we decompose the problem~\eqref{opt2} and introduce two auxiliary variables $\tilde{e}^{u,v}_t = e^{u,v}_t$ and $\lambda^{u,v}_t$ to derive the augmented Lagrangian as
    \begin{equation}
    \begin{aligned}
           \mathcal{L}  = \sum_{u\in\mathbf{U}} 
            \mathcal{C}_{\mathrm{S2}}^{u}(\mathbf{g}^{u}, \mathbf{h}^{u}, \mathbf{c}^{u}, \mathbf{d}^{u}, \mathbf{e}^{u,v})
         &+ \sum_{u\in\mathbf{U}} \sum_{v \in \mathbf{U}} \sum_{t\in\mathbf{H}} 
        \Big[ \frac{\rho}{2} \left( \tilde{e}^{u,v}_t - e^{u,v}_t \right)^{2} \\
         &+ \lambda^{u,v}_t \left( \tilde{e}^{u,v}_t - e^{u,v}_t \right) \Big], \label{lagrangian}
    \end{aligned}    
    \end{equation}
where the parameter $\rho$ is the penalty coefficient and $\tilde{e}^{u,v}_t$ is the auxiliary variable for the energy trading decision of the user $u$. Since $\tilde{e}^{u,v}_t$ must also follow the constraints of the original variable ${e}^{u,v}_t$ in Eq.~\eqref{constraint-trading1}, hence we establish the following constraints:
    \begin{align}
        \tilde{e}_t^{u,v} &= -\tilde{e}_t^{v,u},~\forall t \in \mathbf{H},~\forall u,v \in \mathbf{U}, \label{constraint-auxiliary1} \\
        \tilde{e}_t^{u,v} &= 0,~\textrm{if}~u = v. \label{constraint-auxiliary2}
    \end{align}
    
To shorten the notation, we define the following equivalent variables for $\tilde{e}^{u,v}_t$ and $\lambda^{u,v}_t$
\begin{align}
    &\textrm{Original variables:} \quad & \mathbf{e}^{u} &\triangleq \left\{ {e}^{u,v}_t, \forall v \in \mathbf{U}, \forall t \in \mathbf{H}  \right\}, \\
    &\textrm{Auxiliary variables:} \quad & \tilde{\mathbf{e}}^{u} &\triangleq \left\{ \tilde{e}^{u,v}_t, \forall v \in \mathbf{U}, \forall t \in \mathbf{H}  \right\}, \\
    &\textrm{Dual variables:} \quad & \mathbf{\lambda}^{u} &\triangleq \left\{ \lambda^{u,v}_t, \forall v \in \mathbf{U}, \forall t \in \mathbf{H}  \right\}.
\end{align}

\subsection{Distributed algorithm}
Based on the augmented Lagrangian in Eq.~\eqref{lagrangian}, we split the optimization problem into a set of subproblems. In users' subproblems, the users minimize their own total costs in parallel, given the dual variables $\mathbf{\lambda}^{u}$ and auxiliary variables $\tilde{\mathbf{e}}^{u}$. Then the auxiliary variables and dual variables are updated based on the trading decisions submitted by users denoted as $\mathbf{e}^{u}$. This algorithm iterates to solve the subproblem and update the dual variables until it converges to the optimal solution. 

In the primal subproblem, given the dual variables $\mathbf{\lambda}^{u}$ and auxiliary variables $\tilde{\mathbf{e}}^{u}$, user $u$ solves the following optimization problem:
\begin{align}
    &\min \mathcal{C}_{\mathrm{S2}}^{u}(\mathbf{g}^{u}, \mathbf{h}^{u}, \mathbf{c}^{u}, \mathbf{d}^{u}, \mathbf{e}^{u,v}) {+} \sum_{v \in \mathbf{U}} \sum_{t\in\mathbf{H}} \Big[ \frac{\rho}{2} \left( \tilde{e}^{u,v}_t {-} e^{u,v}_t \right)^{2} {-} \lambda^{u,v}_t e^{u,v}_t \Big] \nonumber\\
    &\quad\quad\quad \quad \mathrm{s.t.} \: \mathrm{constraints} \:\eqref{constraint-load6}, \eqref{constraint-load5},
    \eqref{constraint-load7},
    \eqref{constraint-load1},
    \eqref{constraint-load12} \nonumber\\
    &\quad\quad\quad \quad \mathrm{with} \: \mathrm{variables:} \: \mathbf{g}^{u}, \mathbf{h}^{u}, \mathbf{c}^{u}, \mathbf{d}^{u}, \mathbf{e}^{u,v}. \label{primal}
\end{align}
By solving \eqref{primal}, user $u$ obtains the energy trading decision $\mathbf{e}^{u}$ and sends it to the dual problem for the iteration.
    
We introduce a computing module for the community of users to update the dual variables $\mathbf{\lambda}^{u}$ and auxiliary variables $\tilde{\mathbf{e}}^{u}$ by solving the following problem:
\begin{align}
    &\min \sum_{u\in\mathbf{U}} \sum_{v \in \mathbf{U}} \sum_{t\in\mathbf{H}} \Big[ \frac{\rho}{2} \left( \tilde{e}^{u,v}_t - e^{u,v}_t \right)^{2}
         + \lambda^{u,v}_t \tilde{e}^{u,v}_t  \Big] \nonumber\\
    &\quad\quad\quad \quad \mathrm{s.t.} \: \mathrm{constraints} \:\eqref{constraint-auxiliary1},\eqref{constraint-auxiliary2}
    \nonumber\\
    &\quad\quad\quad \quad \mathrm{with} \: \mathrm{variables:} \: \tilde{\mathbf{e}}^{u}. \label{dual}
\end{align}
The optimal solution to \eqref{dual} is
        \begin{align}
                \tilde{e}^{u,v}_{t+1} = \frac{\rho \left( e^{u,v}_t {-} e^{v,u}_t \right) - \left( \lambda^{u,v}_t - \lambda^{v,u}_t \right) }{2 \rho}, \label{updateenergy}
        \end{align}
and during the algorithm iteration, the dual variable $\lambda^{u,v}_t$ is updated as
    \begin{align}
        \lambda^{u,v}_t \leftarrow \lambda^{u,v}_t + \rho \left( \tilde{e}^{u,v}_t - e^{u,v}_t\right). \label{updatelambda}
    \end{align}
To solve the original optimization problem, we solve problems \eqref{primal} and \eqref{dual} in an iterative manner until they converge. We let the difference between the original variable $\mathbf{e}^{u}$ and its auxiliary variable $\tilde{\mathbf{e}}^{u}$ be one of the convergence measure. The other measure is based on the dual variable $\mathbf{\lambda}(k)$ in the $k$th iteration, and we define $\mathbf{\lambda} \triangleq \left\{ \lambda^{u,v}_t,~\forall u,v,t \right\}$. The solution of the optimization problem converges when this error is small enough, namely 
\begin{equation}
    \sum_{u \in \mathbf{U}} \parallel \mathbf{e}^{u} - \tilde{\mathbf{e}}^{u} \parallel < \epsilon_{1},
    ~ \parallel \mathbf{\lambda}(k) - \mathbf{\lambda}(k-1) \parallel < \epsilon_{2}, \label{eq:converge}
\end{equation}
where $\epsilon_{1}$ and $\epsilon_{2}$ are the convergence thresholds that are usually a very small number. \yq{Eq.~\eqref{eq:converge} specifies the convergence conditions for the distributed algorithm. The convergence conditions are based on two criteria, in which $\sum_{u \in \mathbf{U}} \parallel \mathbf{e}^{u} - \tilde{\mathbf{e}}^{u} \parallel < \epsilon_{1}$ guarantees that the primal variable of energy trading converges, and $\parallel \mathbf{\lambda}(k) - \mathbf{\lambda}(k-1) \parallel < \epsilon_{2}$ ensures the convergence of the dual variable $\mathbf{\lambda}$.}

\begin{algorithm}[!tb]
     \caption{Distributed optimization algorithm}
     \label{alg1} 
     \SetAlgoLined
     \textbf{Initialization}:\\
     iteration index $k {\leftarrow} 1$;\\
     convergence threshold $\epsilon_{1}, \epsilon_{2} {\leftarrow} 0.000001$; \\
     iteration stepsize $\rho(0) {\leftarrow} 1$; \\
     dual variable $\mathbf{\lambda}^{u} {\leftarrow} \mathbf{0}$;
    
    \While{$\sum_{u \in \mathbf{U}} \parallel \mathbf{e}^{u} - \tilde{\mathbf{e}}^{u} \parallel > \epsilon_{1} $ $\lor$ $\parallel \mathbf{\lambda}(k) - \mathbf{\lambda}(k-1) \parallel > \epsilon_{2}$}{
        \For{$u \in \mathbf{U}$}{
        (1) User $u$ obtain auxiliary variable $\tilde{\mathbf{e}}^{u}$ and dual variable $\mathbf{\lambda}^{u}$ from the grid operator;
        
        (2) User $u$ solves primal subproblem Eq.~\eqref{primal};
        
        (3) User $u$ sends the energy-trading decisions $\mathbf{e}^{u}$ to the grid operator;
        }
    
    (4) The grid operator receives $\mathbf{e}^{u}$ from all the users $u \in \mathbf{U}$;
    
    (5) The grid operator computes auxiliary variable $\tilde{\mathbf{e}}^{u}$ using Eq.~\eqref{updateenergy};
    
    (6) The grid operator computes dual variable $\mathbf{\lambda}^{u}$ using Eq.~\eqref{updatelambda};
    
    (7) The grid operator updates $\tilde{\mathbf{e}}^{u}$ and $\mathbf{\lambda}^{u}$ to all the users;
    
    (8) $k \leftarrow k+1$;
    }
 \textbf{Outputs}:\\
 Optimal HAVC management and energy trading schedule $\mathbf{M}^u_{S2} = (\mathbf{g}^{u}, \mathbf{h}^{u}, \mathbf{c}^{u}, \mathbf{d}^{u}, \mathbf{r}^{u}, \mathbf{e}^{u,v})$.
\end{algorithm}

The implementation of the above design is presented as the distributed algorithm illustrated in Fig.~\ref{f3:process}. As shown in the figure, the computing module coordinates the information exchange of $\mathbf{\lambda}^{u}$ and $\tilde{\mathbf{e}}^{u}$ for users. We illustrate the energy trading between a pair of users $u$ and $v$ as an example. The users locally solve Problem~\eqref{primal} to obtain their own energy scheduling decisions including the energy trading decision $\mathbf{e}^{u}$, and update $\mathbf{e}^{u}$ to the computing module. Upon receiving the trading decision $\mathbf{e}^{u}$, the computing module solves Problem~\eqref{dual} and update the auxiliary variables $\tilde{e}^{u,v}_t$ and $\lambda^{u,v}_t$ according to \eqref{updateenergy} and \eqref{updatelambda}. The above two steps iterate multiple times until the solution of the optimization problem converges to the optimal energy schedule $\mathbf{M}^u_{S2}, \forall u \in \mathbf{U}$. Then all the users can execute the optimal energy schedule during the time window $t \in \mathbf{H}$.

Algorithm~\ref{alg1} describes the pseudo-code of the proposed distributed optimization method. Specifically, the algorithm solves Problems~\eqref{primal} and~\eqref{dual} for the original problem~\eqref{opt2} in an iterative fashion. Each user locally solves the internal energy scheduling and external energy trading in \eqref{primal} without sharing any private data. The user only needs to report its trading demand to the computing module. The computing module collects the trading decisions from the users to compute the auxiliary and dual variables in \eqref{updateenergy} and \eqref{updatelambda}. Based on the analysis in \cite{boyd2011distributed}, this distributed algorithm converges to the optimal solution of the original optimization problem~\eqref{opt2}.

\section{Simulation and analysis}\label{sec:eval}

\begin{figure}[!t]
    \centering
    \includegraphics[width=7.5cm]{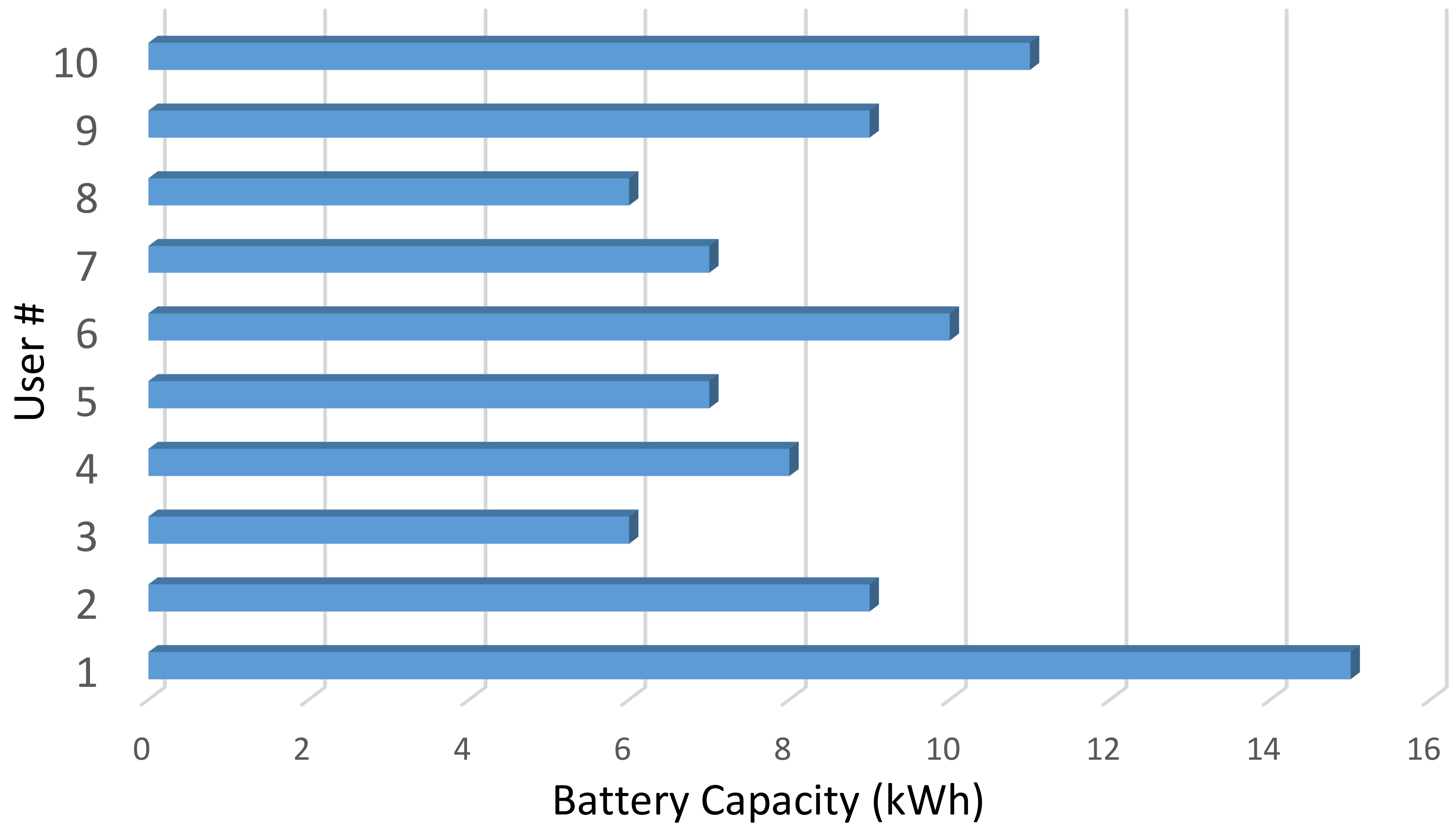}
    \caption{The capacity of the users' battery energy storage systems.}
    \label{f:be}
\end{figure}

\subsection{Simulation setup}\label{subsec:data}
To evaluate the performance of our proposed distributed HVAC management algorithm, we conduct extensive numerical simulation using data collected from real-world homes. We simulate a community of $10$ smart homes and a local grid operator. \yq{The total simulation period is one week (from 2019/9/4 to 2019/9/10) to show the effect of the proposed energy management algorithm.} Each smart home is equipped with HVAC, renewable generators (e.g., solar and wind), battery storage, and other inflexible appliances. The battery capacity and the users' preferences on the indoor temperature are randomly selected from an available set. The battery capacity ranges from 6kWh to 15kWh, as listed in Fig.~\ref{f:be}, and the user's preference ranges from $20^\circ$C to $27^\circ$C.

\begin{figure}[!t]
    \centering
    \includegraphics[width=8.5cm]{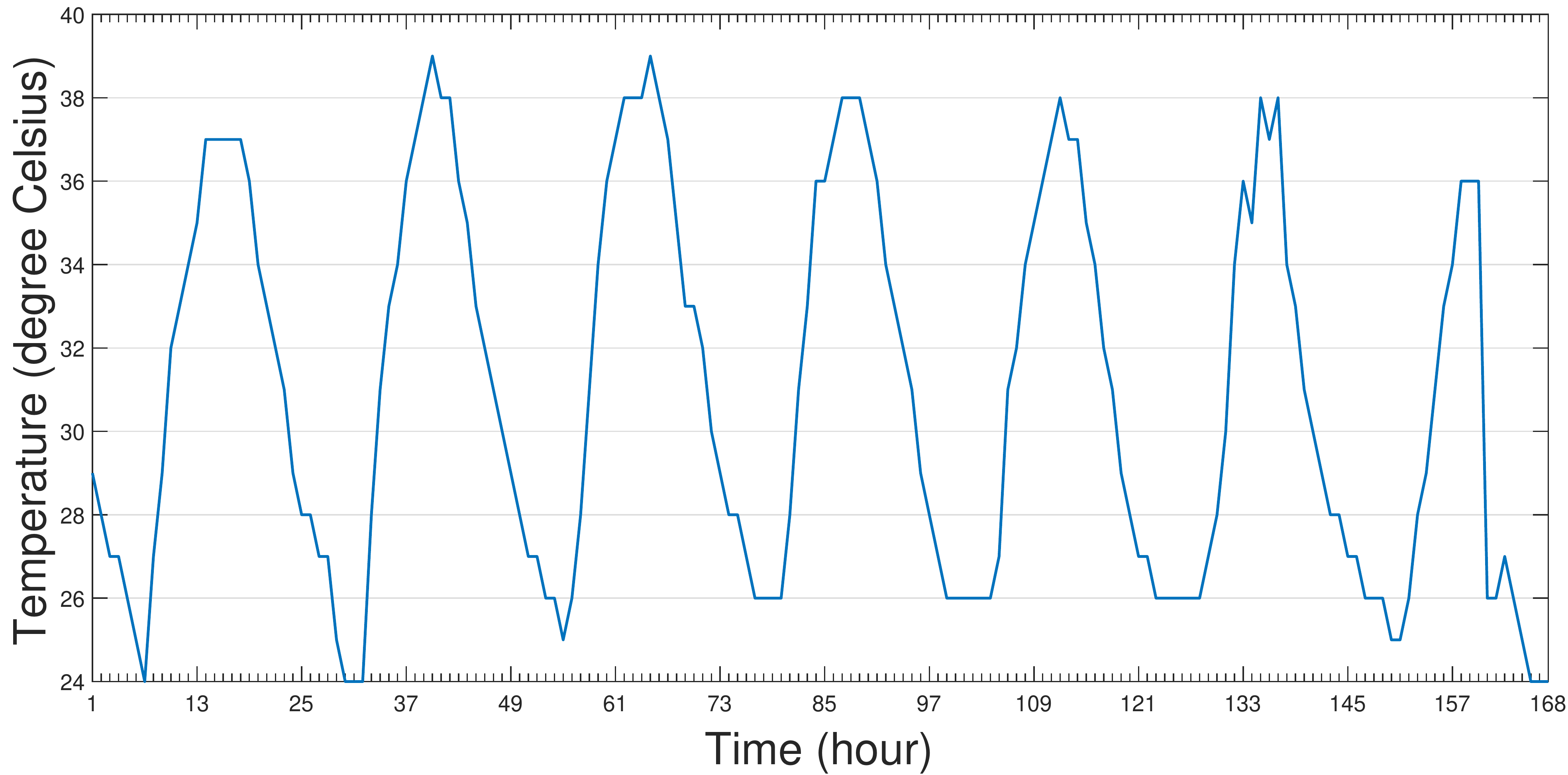}
    \caption{The outdoor temperature of the smart homes during the simulation period of one week.}
    \label{f:tmp}
\end{figure}

\begin{figure}[!tb]
    \centering
    \includegraphics[width=8.5cm]{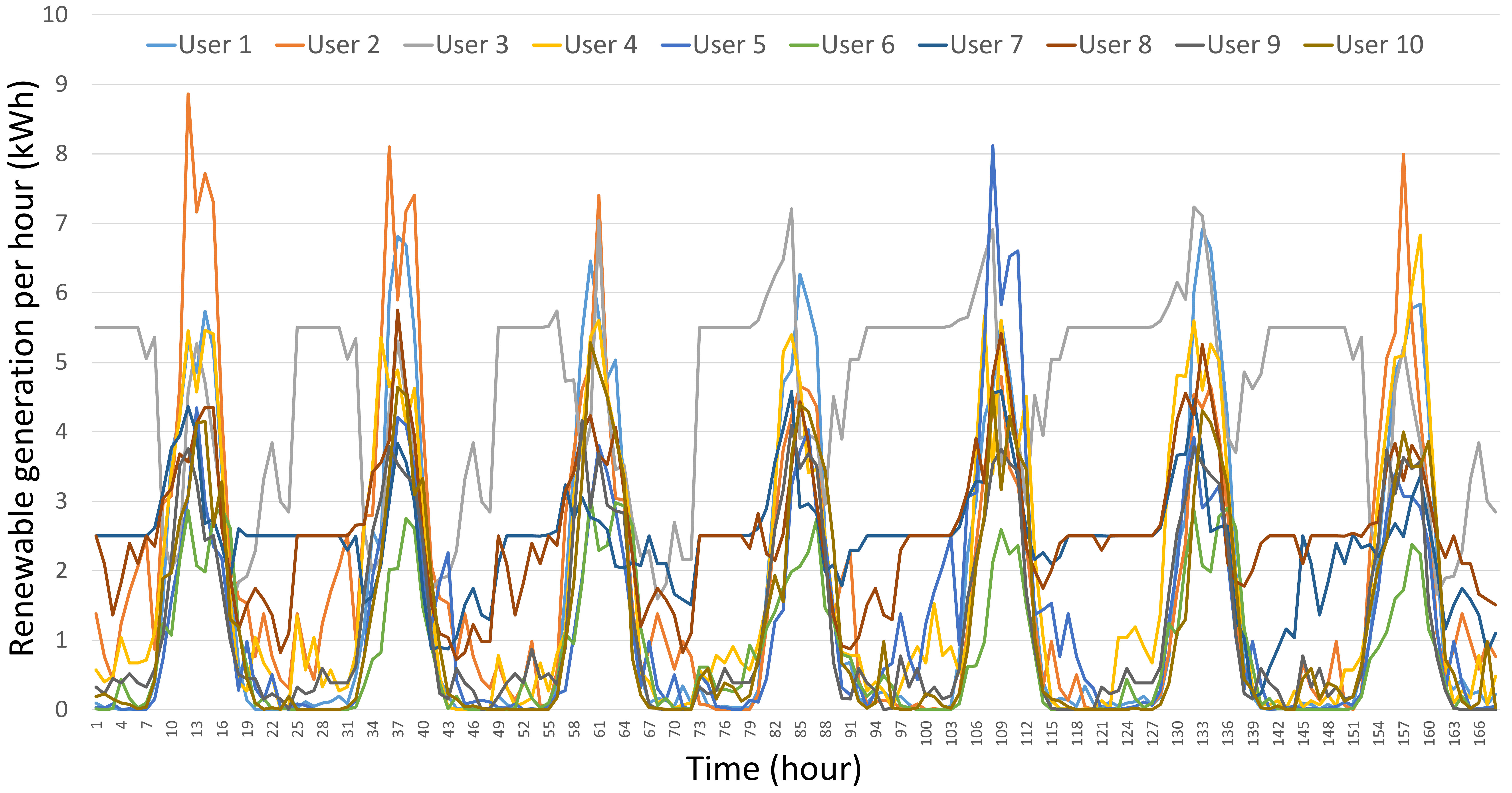}
    \caption{The renewable generation data of the ten users during one week, including the energy generated per hour by both the PV panel and wind turbine.}
    \label{f:re}
\end{figure}

\begin{figure}[!tb]
    \centering
    \includegraphics[width=8.5cm]{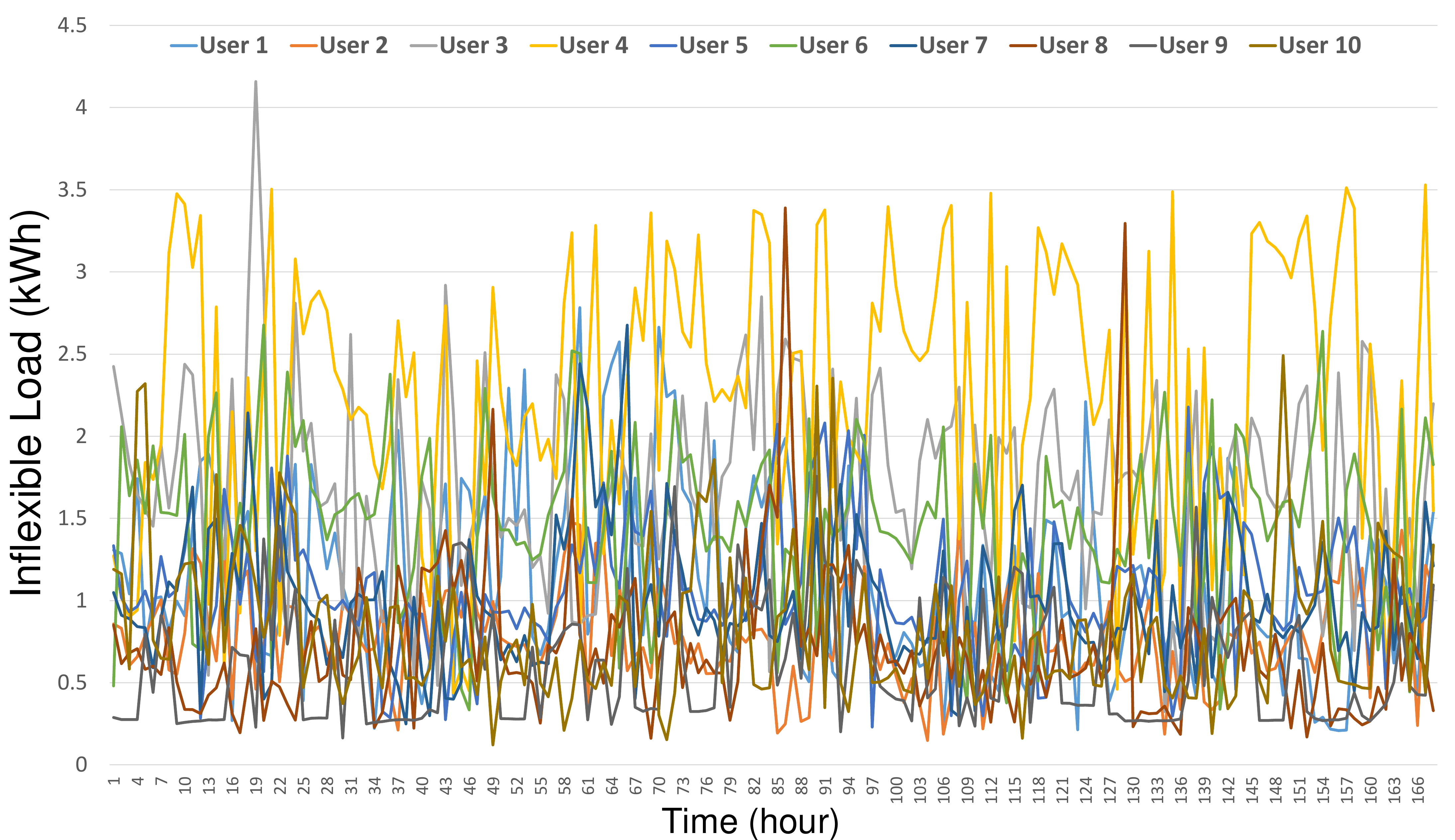}
    \caption{The inflexible load of the ten users during the simulation period of one week (168 hours).}
    \label{f:inflex}
\end{figure}

The simulation data are collected from real-world scenarios, including renewable generation \citep{wang2015joint} and outdoor temperature. The users' renewable generation includes both solar power and wind power over one week \yq{from 2019/9/4 to 2019/9/10} (i.e., 168 hours in total), as shown in Fig.~\ref{f:re}. The outdoor temperature is shown in Fig.~\ref{f:tmp}.
Note that the solar power generation is mainly active during the daytime, and the wind power generation lasts longer and is more diverse. To simulate the inflexible load, we use the power consumption data of the residential household appliances from \cite{pecan}. This dataset includes users' daily electricity usage, such as HVAC and other loads over one week, as shown in Fig.~\ref{f:inflex}. To validate the distributed energy management algorithm developed in Section~\ref{sec:solution}, we simulate a community of 10 users, and each user has \yq{an} HVAC unit and local renewable generation.

\subsection{Algorithm convergence}
First, we validate the convergence performance of the distributed optimization algorithm in Section~\ref{sec:solution} with real-world data. We set convergence thresholds $\epsilon_{1}$ and $\epsilon_{2}$ so that the algorithm iteration ends if the convergence errors are less than the thresholds. The convergence error of the distributed trading algorithms is defined as the sum of the absolute deviation between the actual trading decisions and auxiliary decisions as in Algorithm~\ref{alg1}. In our simulation, we set the convergence thresholds $\epsilon_{1}$ and $\epsilon_{2}$ to be $1 \times 10^{-6}$. We run the distributed optimization algorithm in Matlab on a commodity PC (Intel i7-9700 CPU and 4GB memory), and Algorithm~\ref{alg1} converges within $26$ iterations. The result shows that the proposed distributed trading algorithm converges fast, so the proposed algorithm is feasible in a practical smart grid system.

\subsection{Power scheduling in energy trading scenario} 
We then implement the distributed HVAC management algorithm using the simulation data described in Subsection \ref{subsec:data} to evaluate its performance. Since we solve the day-ahead energy scheduling in this work, the distributed algorithm optimizes the users' hourly energy schedule for the next day. We run the simulations for seven days continuously, so the simulation period is 168 hours. The user's schedule includes renewable energy supply, grid electricity usage, battery operation, and the HVAC energy consumption.

\begin{figure}[!t]
    \centering
    \includegraphics[width=8.5cm]{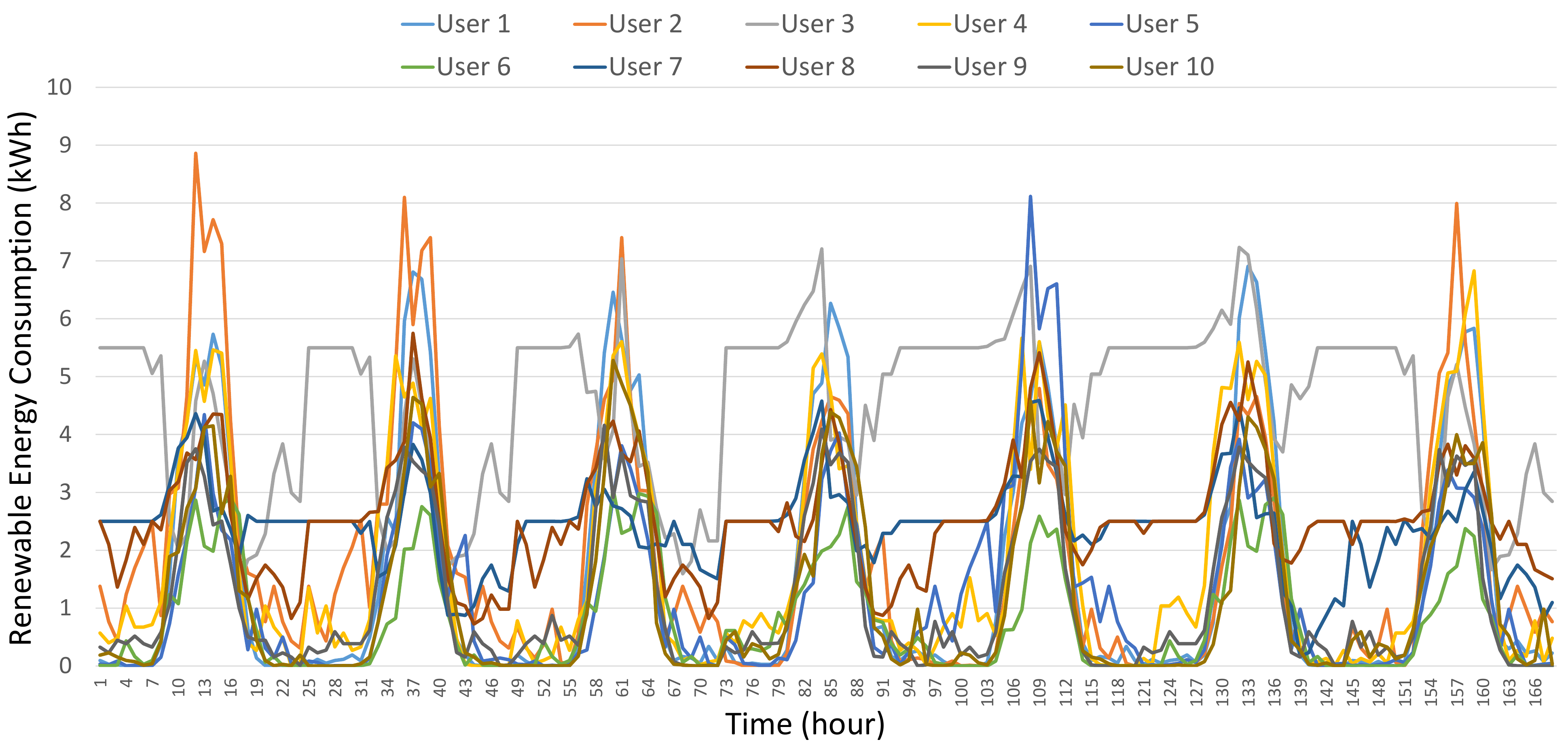}
    \caption{The optimal energy schedule of the renewable energy supply for the ten users during the simulation period. }
    \label{f:rn}
\end{figure}

\begin{figure}[!t]
    \centering
    \includegraphics[width=8.5cm]{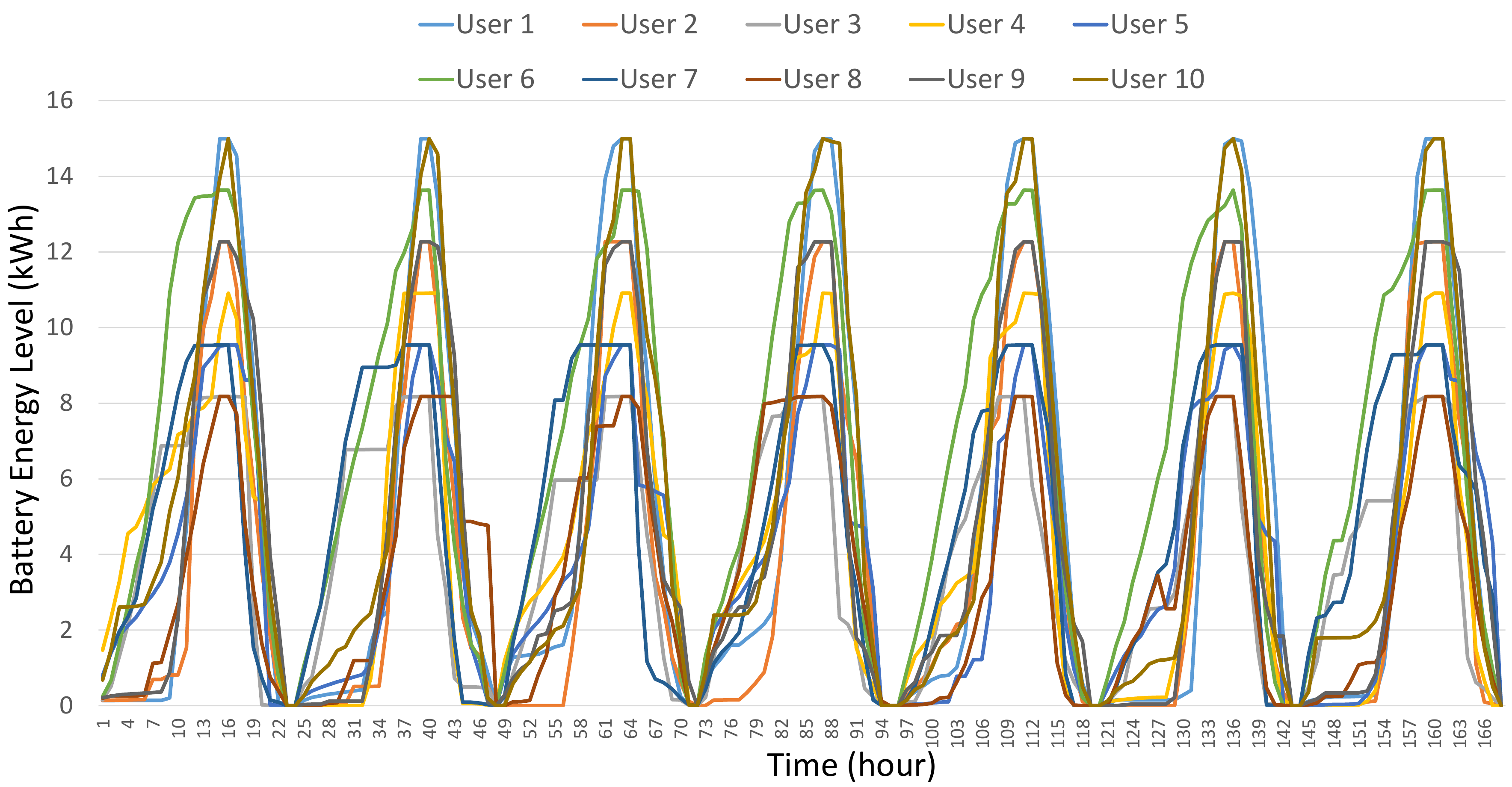}
    \caption{The optimal energy schedule of the battery for the ten users during the simulation period. }
    \label{f:bo}
\end{figure}

\begin{figure}[!tb]
    \centering
    \includegraphics[width=8.5cm]{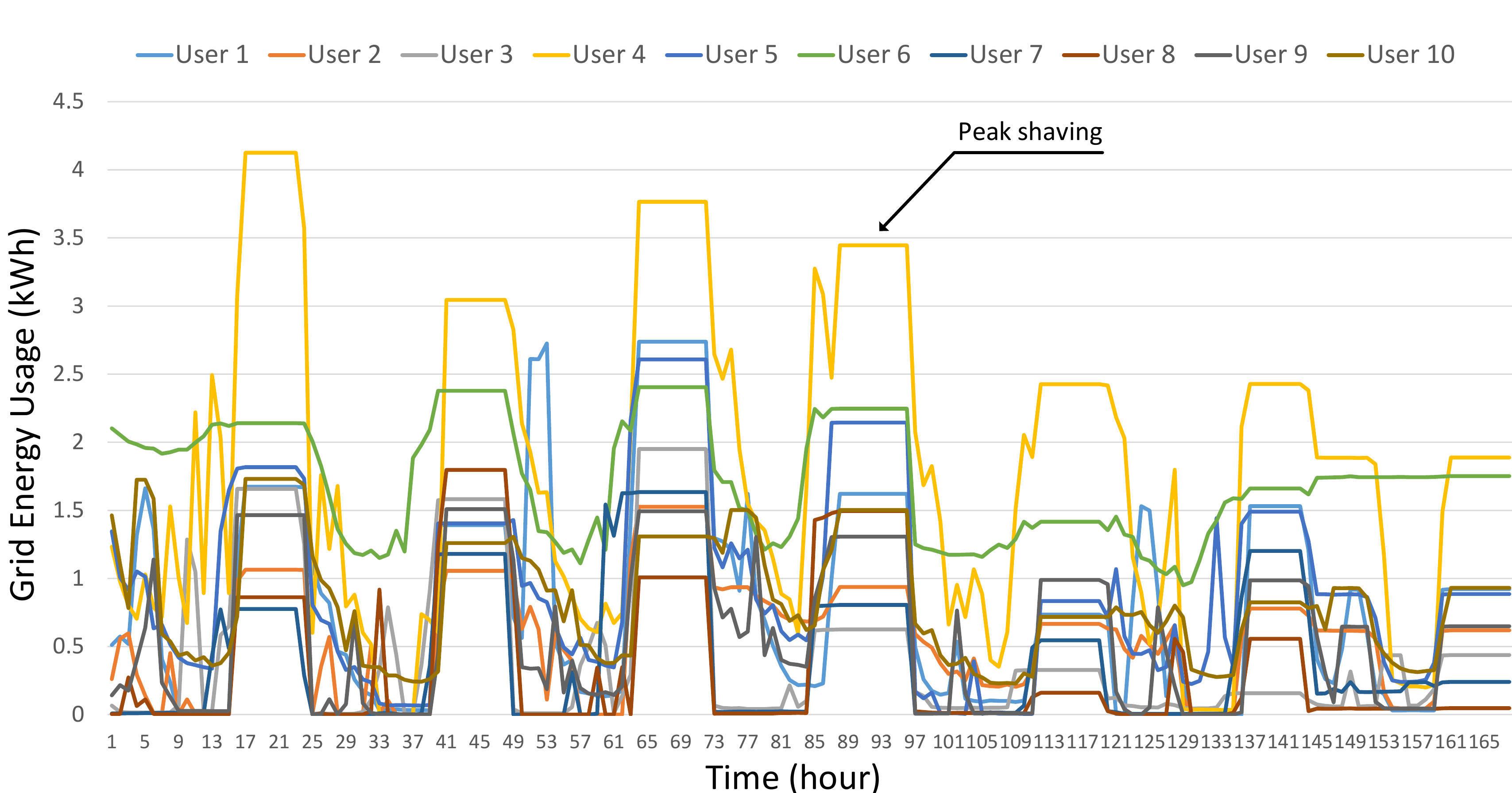}
    \caption{The optimal grid energy schedule for the ten users by the distributed energy management algorithm. }
    \label{f:gu}
\end{figure}
\begin{figure}[!t]
    \centering
    \includegraphics[width=8.5cm]{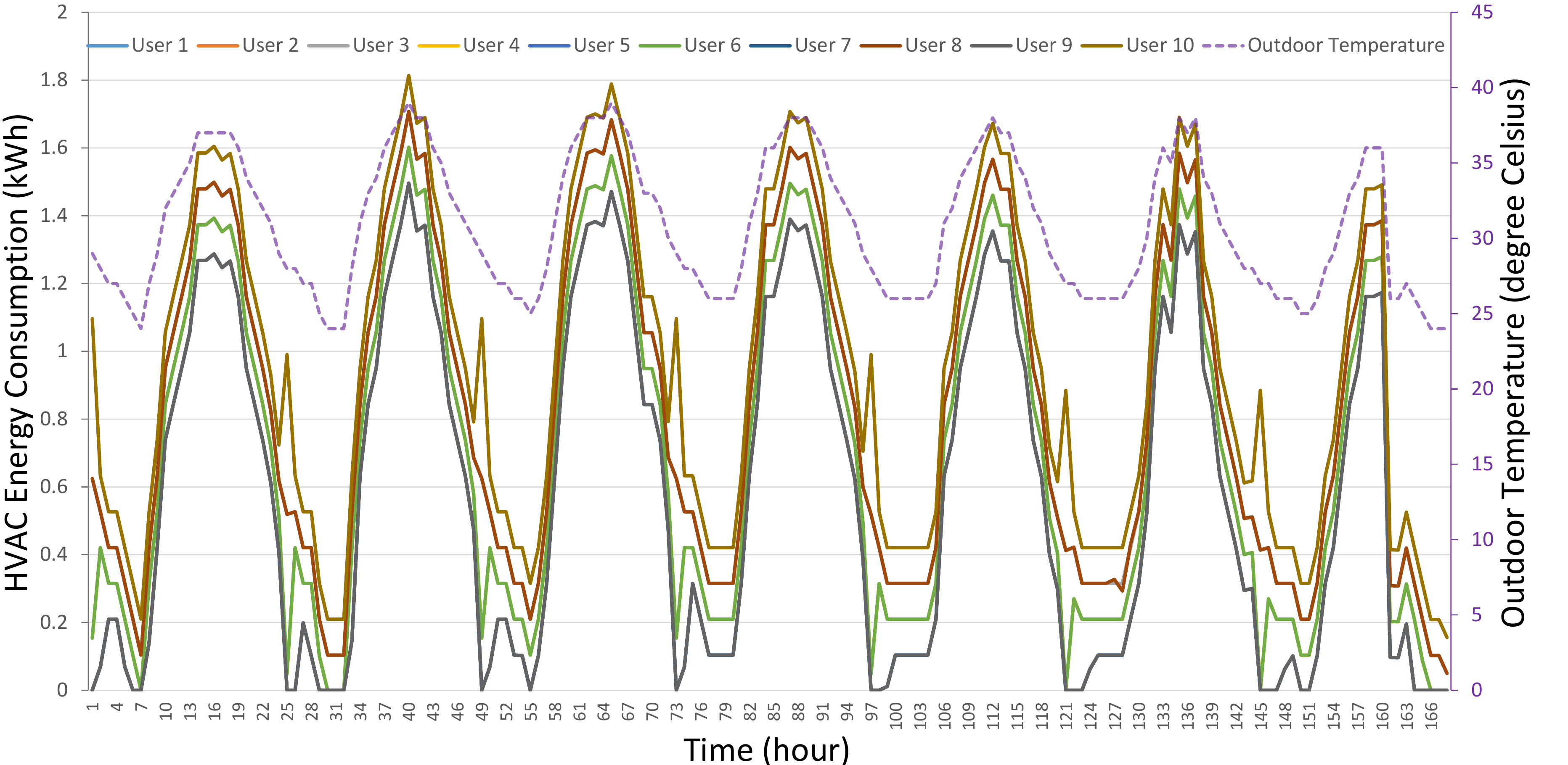}
    \caption{The optimal HVAC energy schedule for the ten users by the distributed energy management algorithm.}
    \label{f:ac}
\end{figure}

Fig.~\ref{f:rn} shows the optimal energy schedule of renewable energy supply for all the users. Comparing Fig.~\ref{f:rn} with Fig.~\ref{f:re}, we see that the users' renewable energy are well utilized through our developed distributed algorithm. The users' battery operation is plotted in Fig.~\ref{f:bo} that shows the energy level of their batteries. We see that the users' batteries store energy during the daytime when the renewable energy generation is adequate, and discharge the energy during the night to serve the loads. Fig.~\ref{f:gu} shows the optimal schedule of the grid energy usage for all the users. Comparing Fig.~\ref{f:gu} with Fig.~\ref{f:rn}, we can see that the grid usage is well complemented with the renewable energy usage, since the distributed algorithm reduces users' dependence on the grid to lower their costs. Moreover, the flat plateaus in Fig.~\ref{f:gu} also demonstrate peak-shaving of the grid supply. The optimal schedule of the HVAC energy consumption with respect to the outdoor temperature is plotted in Fig.~\ref{f:ac}. The figure shows that scheduled HVAC load varies with the temperature changes to keep the indoor temperature around the users' preferred temperatures.

\subsection{P2P energy trading}

\begin{figure}[!t]
    \centering
    \includegraphics[width=8.5cm]{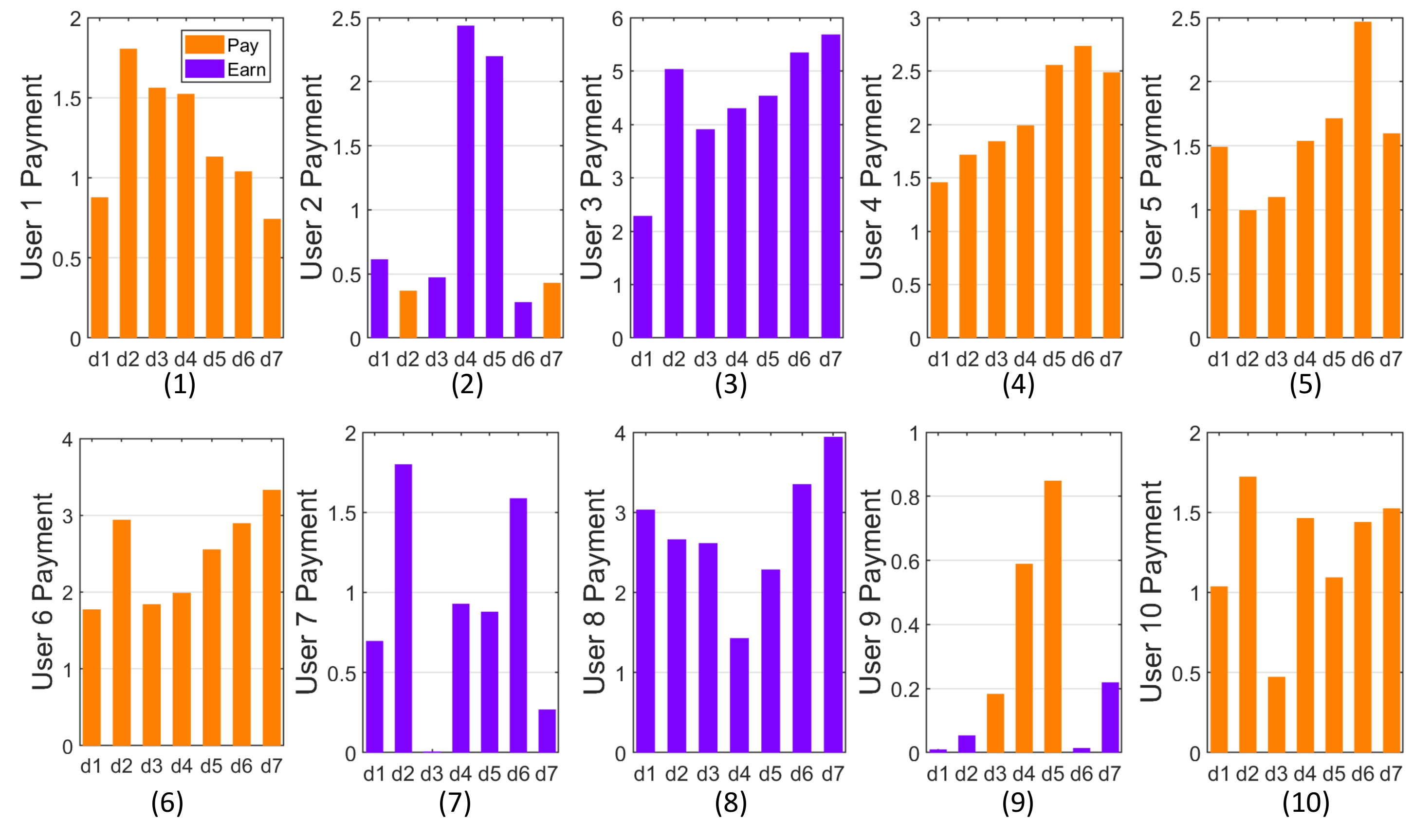}
    \caption{The users' payments for the P2P energy trading during the simulation period of one week (d1 to d7). The payment is the gross sum of the user's payment and profit on that day.}
    \label{f:pay}
\end{figure}
During the simulation, the users actively trade energy with each other to minimize their energy costs. We plot the payments for P2P energy trading of all the 10 users over one week in Fig.~\ref{f:pay}. Since the payments are settled at the end of every trading day, the figure is plotted with one-day granularity. Based on the payment results, 10 users can be classified into three categories. Some users (such as users 1, 4, 5, 6) pay energy-trading fees to other users during the simulated week, because they are short of local renewable supply and buy energy from other users frequently. Some users (such as users 3 and 8) earn extra profits every day since they have extra energy to sell. The rest of the users pay for energy trading on some days and also earn profits on the other days, depending on the amount of renewable energy generated on that day. The above results demonstrate that the platform provides great flexibility to users to use energy more efficiently, and they all benefit from the trading despite their different patterns.

\begin{figure}[!t]
    \centering
    \includegraphics[width=8.5cm]{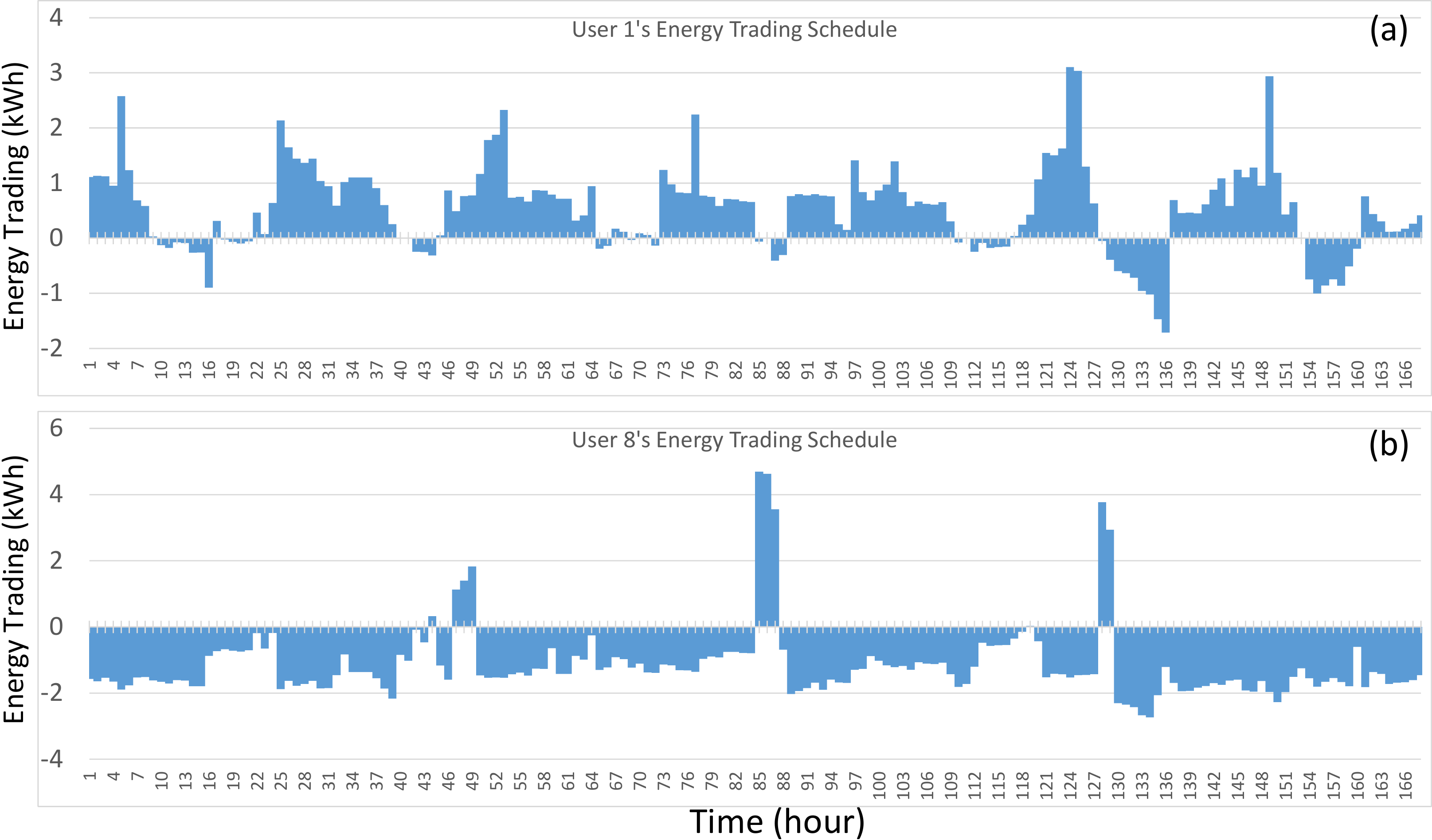}
    \caption{The optimal energy trading schedule of two typical users (a) user 1 and (b) user 8). Note that positive values indicate buying energy and negative values indicate selling energy.}
    \label{f:et}
\end{figure}

To show the details of the energy trading process, we plot the energy trading operation of two typical users (user 1 and user 8) in Fig.~\ref{f:et}. These two users have different trading patterns. User 1 is short of renewable energy supply in the daytime and often purchases energy from other users.
By contrast, user 8 is energy self-sustained and thus sells a lot of energy to other users. Fig.~\ref{f:et} shows active energy trading among users, where positive values denote energy purchase and negative values denote energy selling. The results demonstrate that our distributed optimization algorithm well incentivizes the users to participate in the distributed HVAC management platform for reducing the users' costs.

\subsection{Cost analysis}

\begin{figure}[!t]
    \centering
    \includegraphics[width=8.5cm]{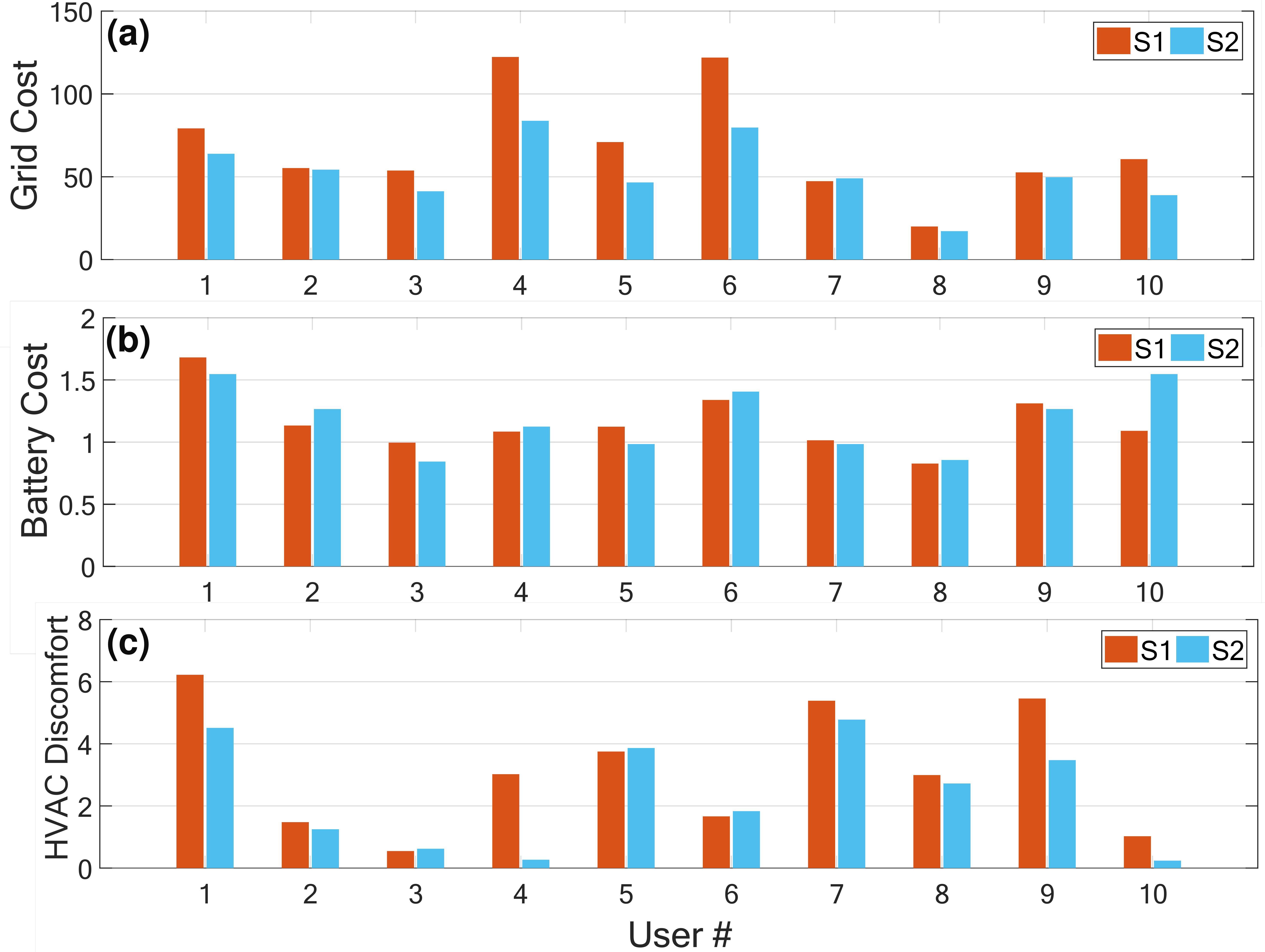}
    \caption{The user's costs on (a) grid electricity, (b) battery degradation, and (c) HVAC discomfort during the simulation period. Scenario 1 (S1) is the standalone HVAC mode without energy trading. Scenario 2 (S2) is the distributed HVAC management with energy trading.}
    \label{f:detailcost}
\end{figure}

\yq{We elaborate the detailed costs of grid electricity payment, battery degradation, and HVAC discomfort in Fig.~\ref{f:detailcost}. We can see that energy trading reduces all the users' dependency on the grid and reduce the grid energy cost for users. Also, all the users reduce the discomfort cost when using the HVAC, implying that the energy trading among smart homes can help users to schedule their energy consumption and improve their indoor comfort. Due to the diverse renewable energy and consumption profiles, different users behave differently in energy trading. To assist the energy trading and schedule their in-home energy consumption, energy storage is used to meet individual needs of charging and discharging, leading to different battery operating patterns. So we see that some users (such as users \#2, \#4, \#6, \#10) have a higher battery cost, and the rest of users have a lower battery cost.}

\subsection{Total cost reduction}
\begin{table}[!tb]
    \centering
    \footnotesize
    \renewcommand{\arraystretch}{1.1}
    \caption{The comparison of all users' costs in the two scenarios.}
    \label{t1:cost}
    \begin{tabular}{c c c c}
        \hline
        \multirow{2}{*}{User} & \multicolumn{2}{c}{Cost} & \multirow{2}{*}{Reduction (\%)} \\
        \cline{2-3}
        & Scenario 1 & Scenario 2 &   \\
        \hline
        1 & 84.2 & 70.7 &  16\% \\
        2 & 56.6 & 44.5 &  21\% \\
        3 & 54.9 & 31.5 &  42\% \\
        4 & 123.5 & 106.5 &  14\% \\
        5 & 75.4 &  58.7 &  22\% \\
        6 & 123.4 &  103.4 &  16\% \\
        7 & 51.7 & 38.6 &  25\% \\
        8 & 21.1 & 3.5 &  83\% \\
        9 & 57.3 & 45.3 &  21\% \\
        10 & 61.9 & 47.5 &  23\% \\
        \hline
    \end{tabular}
\end{table}
\yq{
\begin{figure}[!t]
    \centering
    \includegraphics[width=8.5cm]{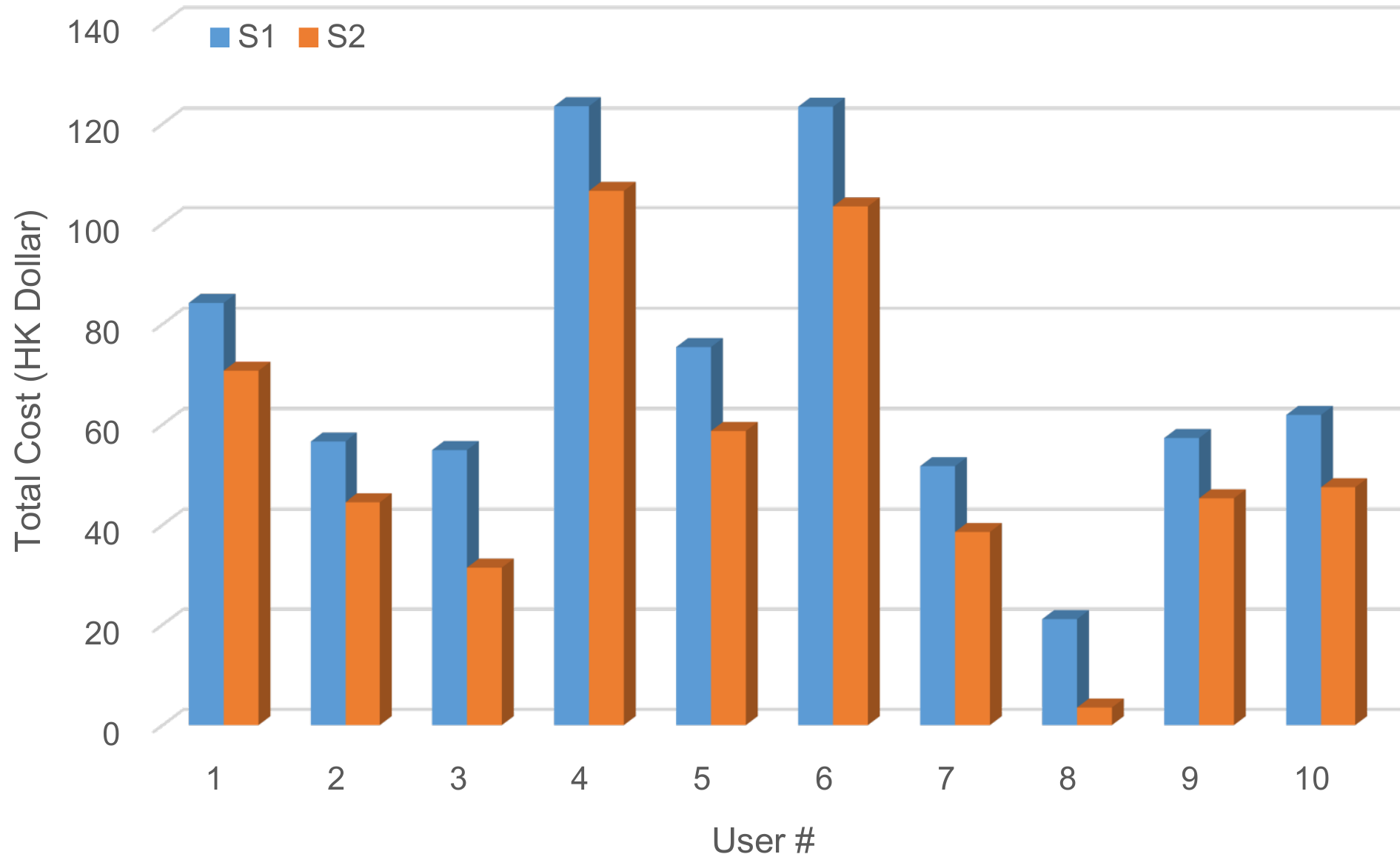}
    \caption{The comparison of the users' costs in the two scenarios. Scenario 1 (S1) is the standalone HVAC mode without energy trading. Scenario 2 (S2) is the distributed HVAC management with energy trading. Date: from 2019/9/4 to 2019/9/10 (Summer).}
    \label{f:overall_cost}
\end{figure}}
To evaluate the performance of the distributed energy management platform, we compare each user's cost in the two scenarios defined in Section~\ref{sec:formulation}. As depicted in Fig.~\ref{f:overall_cost}, the blue bars indicate the costs of all users in Scenario 1, where the users independently manage their energy usage without energy trading. The orange bars indicate the users' costs in Scenario 2, where the users employ energy trading and the proposed distributed energy management algorithm. Comparing the costs in both scenarios, we see that all the users reduce their costs through the distributed energy management and energy trading. By employing the distributed energy management algorithm, the cost reduction of each user is listed in Tab.~\ref{t1:cost}. The average cost reduction of all the users is 23\%.

\section{Conclusions and future works}\label{sec:conclusion}
\yq{This work presented a novel HVAC management method that improves the energy efficiency of smart homes with battery energy storage systems (BESS) and transactive energy. Specifically, we designed a distributed optimization algorithm based on the ADMM method to optimize the users' energy usage as well as facilitate P2P energy trading. Unlike the existing centralized algorithm, the proposed distributed algorithm is privacy-preserving because it does not disclose the users' energy usage while converges to the optimal solution. Furthermore, we proved that this approach is practical and effective with extensive simulations using real-world data. The results showed that the proposed method can utilize the battery system and transactive energy to effectively reduce the costs of all the users.}

\yq{During the study of this paper, we found three issues that are worth further investigation as our future work. First, we will explore reducing the computational complexity of the distributed algorithm to make it converge fast even with a large number of users. Second, we aim to combine the proposed HVAC management method with blockchain technology to build a trusted and decentralized energy management system. Third, one limitation of this work is that we did not model the power loss. We will consider the power loss of the distribution network and the payback period of hardware and software upgrades for energy trading in our future work.}

\section{ACKNOWLEDGMENTS}
This work is in part supported by the National Natural Science Foundation of China (project 61901280) and the FIT Academic Staff Funding of Monash University.

\begin{figure}[!htb]
    \centering
    \includegraphics[width=7.5cm]{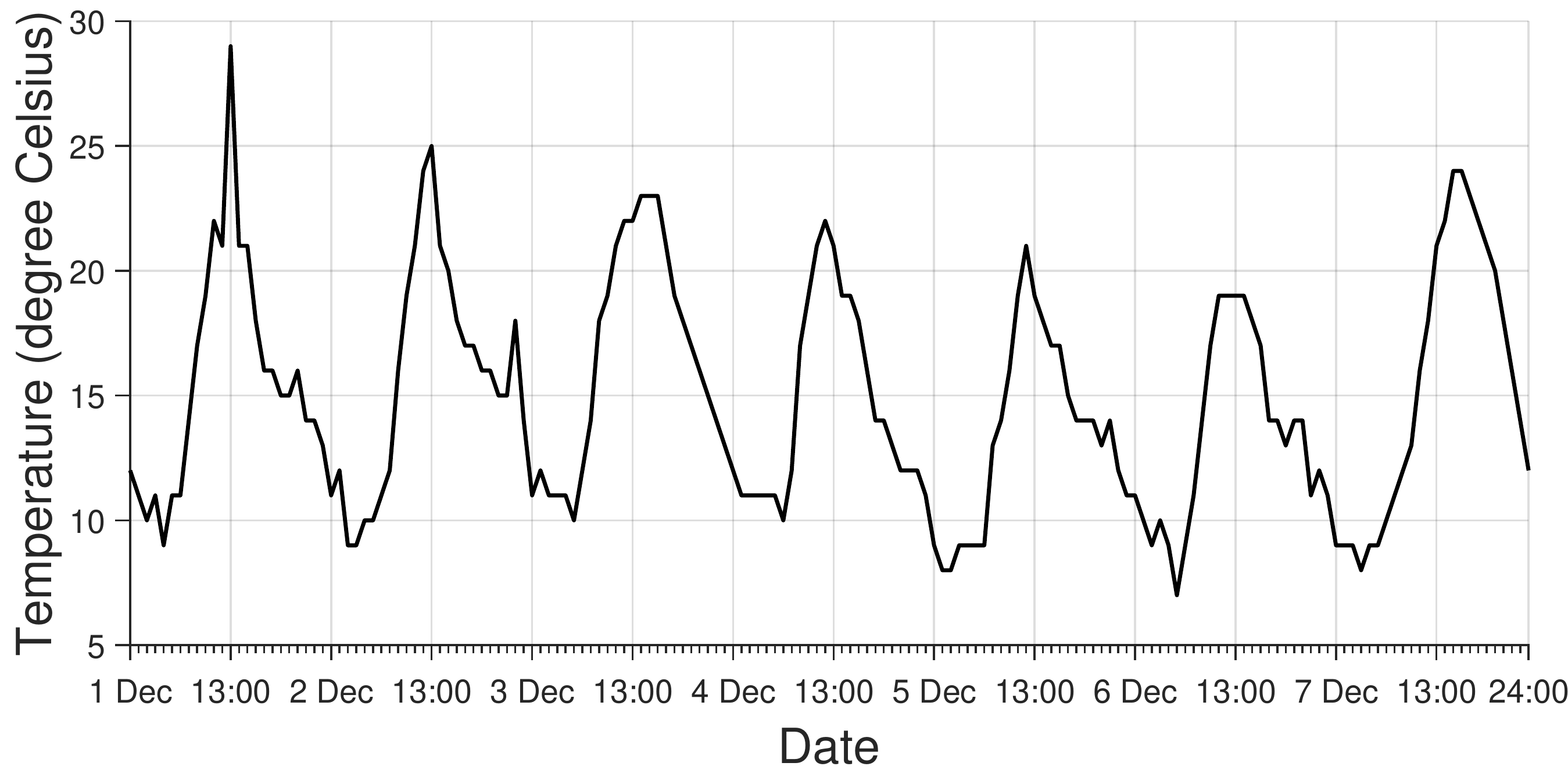}
    \caption{The outdoor temperature of the Los Angeles area during the simulation period of one week. Date: from 2020/12/1 to 2020/12/7 (winter).}
    \label{f:tmp_win}
\end{figure}

\begin{figure}[!htb]
    \centering
    \includegraphics[width=8.5cm]{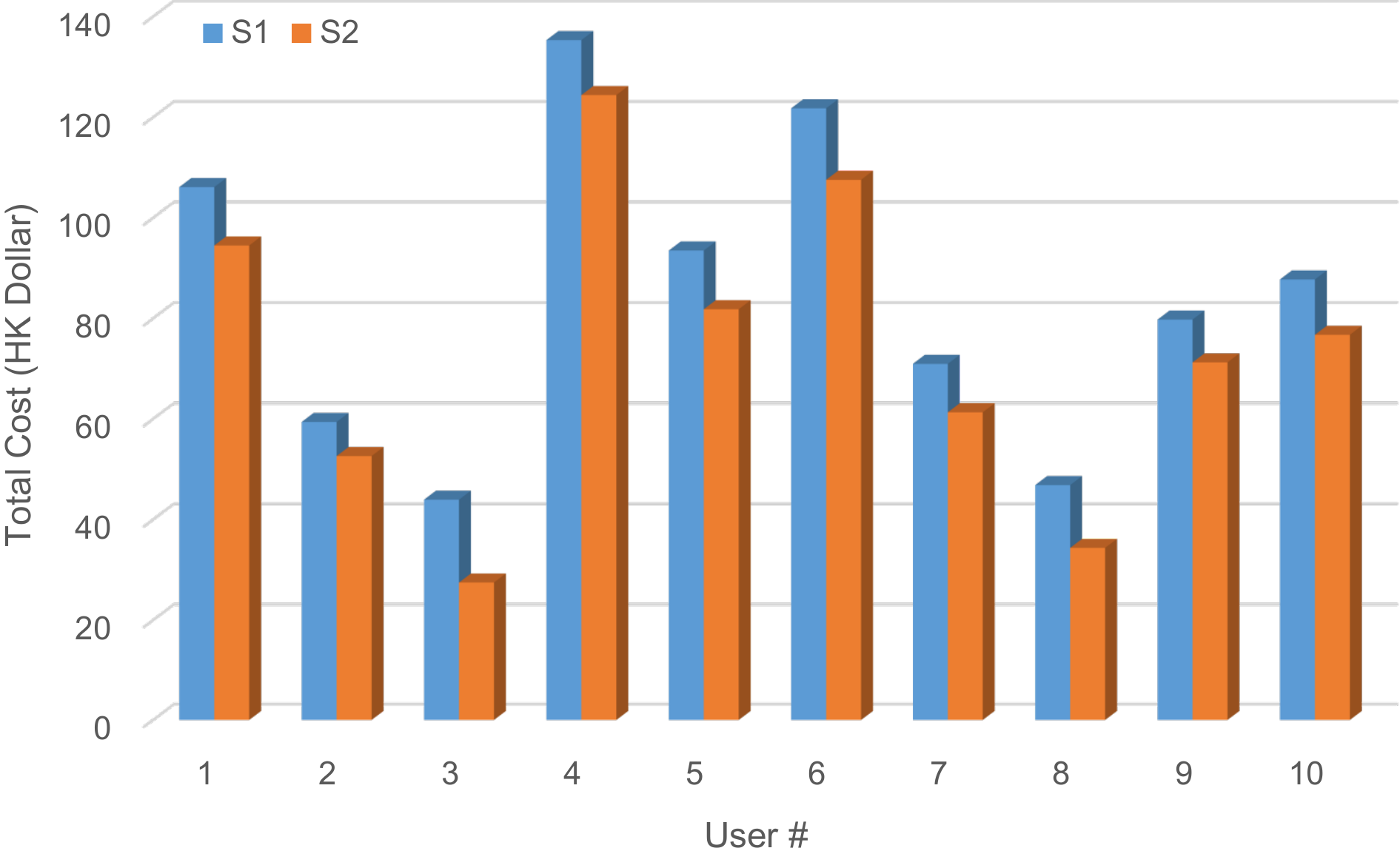}
    \caption{The comparison of the users' costs in the two scenarios. Scenario 1 (S1) is the standalone HVAC mode without energy trading. Scenario 2 (S2) is the distributed HVAC management with energy trading. Date: from 2020/12/1 to 2020/12/7 (winter).}
    \label{f:cost_w}
\end{figure}

\appendix
\yq{\section{Simulation results during the winter period}
In the main text, the outdoor temperature data used in the simulation is collected in the summer period from 2019/9/4 to 2019/9/10. To evaluate the performance of the proposed algorithm, we conduct more simulations under the outdoor temperature in winter. We use the outdoor temperature of Los Angeles during the period of 2020/12/1-2020/12/7 as shown in Fig.~\ref{f:tmp_win}. The simulation results in winter are close to that in summer, but the reduction of total cost in winter (13.4\%) is less than that in summer (23\%). We plot the total cost of all the users in Fig.~\ref{f:cost_w}.}

\begin{figure*}[!htb]
    \centering
    \includegraphics[width=16cm]{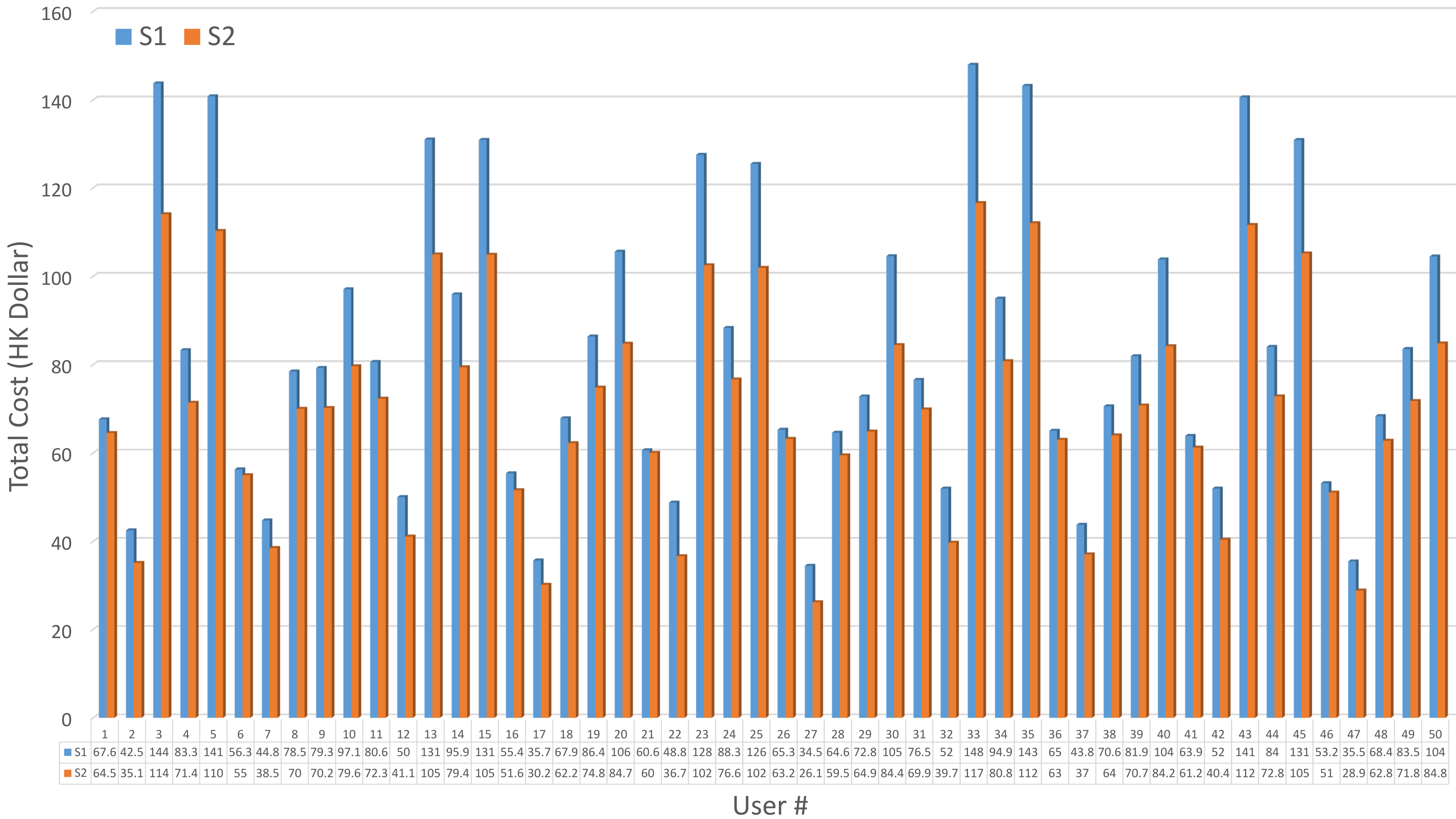}
    \caption{The comparison of the users' costs in the two scenarios. Scenario 1 (S1) is the standalone HVAC mode without energy trading. Scenario 2 (S2) is the distributed HVAC management with energy trading. Date: from 2020/12/1 to 2020/12/7 (winter).}
    \label{f:u50}
\end{figure*}

\yq{\section{Simulation with a larger group of users}
In the main text, we only simulate $10$ users as an example to validate our proposed algorithm's feasibility and effectiveness. However, $10$ users are too few for a practical smart grid system. To evaluate the proposed algorithm's performance on a smart grid of a larger size, we conduct the simulation with $50$ users, using the same simulation setup as our previous simulation. As shown in Fig.~\ref{f:u50}, the users' total costs are also effectively reduced, and the system's overall cost is reduced by 15.15\%, which validates that the proposed algorithm is applicable in larger smart grid systems.
}

\bibliography{mybibfile}

\end{document}